\documentclass[fleqn,final,5p,times]{elsarticle}

\usepackage{amssymb}
\usepackage{amsthm}

\usepackage{amsmath}
\usepackage{booktabs}
\usepackage{calc}
\usepackage{natbib}
\usepackage{bm}
\usepackage{siunitx}
\usepackage{xfrac}
\usepackage{nth}
\usepackage{makecell}
\usepackage{textcomp}
\usepackage{notoccite}

\usepackage{algorithm}
\usepackage{algpseudocode}
\usepackage{algorithmicx}
\algblock[<block>]{<start>}{<end>}
\usepackage{tikz}
\usetikzlibrary{colorbrewer}
\usetikzlibrary{positioning}
\usetikzlibrary{shapes.geometric}
\usetikzlibrary{arrows}
\usetikzlibrary{external}

\usepackage{pgfplots}
\usepgfplotslibrary{groupplots}
\usepgfplotslibrary{colorbrewer}
\usepgfplotslibrary{fillbetween}

\pgfplotsset{
  compat=1.16,
  cycle list/Paired,
  grid=both,
  grid style={line width=.1pt, draw=gray!5},
  major grid style={line width=.2pt,draw=gray!25},
  siunitxlabels/.style={
    /pgfplots/typeset ticklabel/.code={{\pgfmathparse{\tick}$\num[zero-decimal-to-integer]{\pgfmathresult}$}},
  },
}

\newcommand{\cred}{     230,  25,  75}
\newcommand{\cgreen}{    60, 180,  75}

\newcommand{\cblue}{      0, 130, 200}
\newcommand{\corange}{  245, 130,  48}
\newcommand{\ccyan}{     70, 240, 240}
\newcommand{\cmagenta}{ 240,  50, 230}

\newcommand{\cteal}{      0, 128, 128}
\newcommand{\clavender}{220, 190, 255}
\newcommand{\cbrown}{   170, 110,  40}
\newcommand{\cbeige}{   255, 250, 200}
\newcommand{\cmaroon}{  128,   0,   0}

\newcommand{\cnavy}{      0,   0, 128}
\newcommand{\cgrey}{    128, 128, 128}

\definecolor{color0}{RGB}{\cmaroon}
\definecolor{color1}{RGB}{\corange}
\definecolor{color2}{RGB}{\cteal}
\definecolor{color3}{RGB}{\ccyan}
\definecolor{color4}{RGB}{\cgrey}
\definecolor{color5}{RGB}{\cnavy}
\definecolor{color6}{RGB}{\cred}
\definecolor{color7}{RGB}{\cbrown}
\definecolor{color8}{RGB}{\cmagenta}
\definecolor{color9}{RGB}{\cgreen}
\definecolor{color10}{RGB}{\cbeige}
\definecolor{color11}{RGB}{\cblue}
\definecolor{color12}{RGB}{\clavender}

\definecolor{colora}{RGB}{\cblue}
\definecolor{colorb}{RGB}{\cbrown}
\definecolor{colorc}{RGB}{\cgreen}
\definecolor{colord}{RGB}{\cred}
\definecolor{colore}{RGB}{\corange}
\definecolor{colorf}{RGB}{\cteal}

\newcommand{\ppvec}[1]{\bm{#1}}
\newcommand{\code}[1]{\texttt{#1}}
\newcommand{\ppRe}{\ensuremath{\mathrm{Re}}}
\newcommand{\ppMa}{\ensuremath{\mathrm{Ma}}}

\usepackage{xcolor}

\usepackage[hidelinks]{hyperref}

\usepackage{tabularx}
\usepackage{multirow}
\newcolumntype{"}{@{\hskip\tabcolsep\vrule width 1pt\hskip\tabcolsep}}
\usepackage{subcaption}
\usepackage{nicefrac}
\usepackage[capitalise]{cleveref}
\usepackage{xspace}

\newcommand{\flexi}{FLEXI\xspace}
\newcommand{\galaexi}{GA\-L{\AE}\-XI\xspace}

\newcommand{\numThreePlaces}[1]{\num[round-mode=places,round-precision=3]{#1}}

\newcommand{\testfunc}{\boldsymbol{\phi}}
\newcommand{\fphysref}{\boldsymbol{\mathcal{F}}}

\newcommand{\cons}{\mathbf{q}}

\newcommand{\lr}[1]{\left( #1 \right)}
\newcommand{\avg}[1]{\ensuremath{\{\!\!\{#1\}\!\!\}} }

\usepackage{listings}
\journal{}

\begin{document}

\begin{frontmatter}

\title{An Architecture-Agnostic High-Order Discontinuous Galerkin Framework for Compressible Flows}

\author[label1]{Spencer Starr\corref{cor1}}
\ead{spencer.starr@iag.uni-stuttgart.de}

\author[label1]{Yannik Feldner}

\author[label1]{Patrick Kopper}

\author[label1]{Marcel Blind}

\author[label1]{Daniel Kempf}

\author[label1]{Jannik Schrempp}

\author[label1]{Felix Rodach}

\author[label1]{Andrea Beck}

\author[label1]{Anna Schwarz}

\address[label1]{Institute of Aerodynamics and Gas Dynamics, University of Stuttgart, Wankelstra{\ss}e 3, 70563 Stuttgart, Germany}

\cortext[cor1]{Corresponding author}

\begin{abstract}
  With the recent proliferation of heterogeneous, GPU-accelerated supercomputers, high-order computational fluid dynamics (CFD) simulations of complex, turbulent flows are more accessible than ever. To leverage the computing power of these machines, CFD software must adapt. However, complicating the situation is the emerging need to support hardware from multiple GPU vendors. Addressing this need is the GPU-accelerated, discontinuous Galerkin spectral element method (DGSEM) framework \galaexi, a high-order, open source, architecture-agnostic toolchain for the study of complex, compressible, turbulent flows on unstructured, hexahedral grids. GPU-accelerated computations with \galaexi are possible on GPU hardware by interfacing Fortran source code to the vendor models CUDA C++ for NVIDIA and HIP C++ for AMD.

  The DGSEM implementation in \galaexi was verified using the method of manufactured solutions to rigorously confirm the
  expected order of convergence. Simulations of a compressible Taylor-Green-Vortex also demonstrated excellent agreement with reference solutions across all supported architectures. \galaexi achieved near ideal strong and weak scaling on GPU hardware from both NVIDIA and AMD. In the largest case, \galaexi performed a simulation with 67.1 billion degrees of freedom on \num{65536} AMD MI250X graphics compute devices with a parallel efficiency of 82.6\%. Comparing node-to-node performance, GPU simulations offered speedups between 7.75x and 8.08x over CPU computations in time-to-solution while consuming less than half the energy. To demonstrate {\galaexi}'s effectiveness for production-scale simulations, wall-resolved large eddy simulations of the transonic flow past a NACA 64A-110 airfoil and an ONERA OAT15A airfoil under shock buffet conditions were computed.
\end{abstract}


\end{frontmatter}

\section{Introduction}
\label{sec:introduction}

In the ever-evolving landscape of computational fluid dynamics~(CFD), there is a constant push toward larger and more complex problems. One discipline of CFD where this is especially true is the study of three-dimensional, compressible, turbulent flows. This regime contains large disparities in temporal and physical scales along with elaborate interactions between turbulent structures and other flows features like shock waves. The situation is complicated even further when such flows involve complex, detailed geometries, which is often the case in applications such as supercritical wing aerodynamics or turbomachinery.

State-of-the-art research in this area frequently uses\\ approaches such as direct numerical simulation (DNS) or large-eddy simulation (LES) which, for complex geometries, typically require high-order numerical approximations and high-fidelity, unstructured meshes. Such simulations demand a significant computational cost, requiring the use of supercomputers for their calculation to be possible. Since the beginning of high performance computing~(HPC), computational capacity of supercomputers has steadily increased over time, accommodating studies of turbulent flow fields of simultaneously increasing fidelity and complexity.

Currently, the largest computers in the world are capable of performing more than $10^{18}$ floating point operations per second (FLOPs)~\cite{top500nov25}, or one exaFLOP. To achieve those groundbreaking computational capacities, hardware manufacturers have turned to specialized computer chips, generally named accelerators, that offer greater computing capability over traditional CPUs. The most popular type of accelerator is the graphics processing unit~(GPU), whose hardware is specifically suited for the rendering of computer graphics. In the last decade, their use in general HPC applications has increased significantly.

While the number of supercomputers featuring GPUs has risen, the number of manufacturers in the market has also risen. Looking back five years, of the 12 GPU accelerated systems in the top 25, every one used NVIDIA GPUs~\cite{top500nov20}. Now, only four of the systems in the top ten of the current Top500~\cite{top500nov25} utilize GPUs from NVIDIA, while a further four feature AMD GPUs and a ninth system uses GPUs from Intel. For CFD software developers, supporting computations on GPU hardware from multiple vendors is quickly becoming a requirement. Due to competing, manufacturer-specific programming models for GPUs, this requirement presents a significant roadblock to producing CFD software that can leverage all current HPC systems.

\begin{table*}[t]
  \centering
  \caption{
    Comparison of multi-architecture CFD solvers intended for high-order solutions of three-dimensional, compressible flows on unstructured grids.
  }
  \label{tab:codes}
  \begin{tabular}{|p{2cm}||p{1.5cm}|p{3cm}|p{2cm}|p{3cm}|p{2.5cm}|}
    \hline
    \thead{Software} & \thead{Numerical\\Method} & \thead{Shock\\Capturing} & \thead{Programming\\Language} & \thead{Porting\\Approach} & \thead{Supported\\GPU Vendors} \\
    \hline
    \hline
    \makecell*[l]{SOD2D~\cite{gasparino2024sod2d}} & SEM & \makecell*[l]{Entropy Viscosity\\ Stabilization} & Fortran & OpenACC & \makecell{NVIDIA, AMD} \\
    \hline
    \makecell*[l]{PyFR~\cite{witherden2025pyfr}} & FR & Entropy Filtering & Python & \makecell*[l]{Mako templates /\\ code generation} & \makecell*[l]{NVIDIA, AMD,\\ Intel, Apple Metal} \\
    \hline
    \makecell*[l]{\galaexi} & DGSEM & FV sub-cell & Fortran & C++ vendor models (CUDA, HIP) & \makecell{NVIDIA, AMD} \\
    \hline
  \end{tabular}
\end{table*}

Several software frameworks have emerged that overcome the portability barrier and enable architecture-agnostic solutions for complex flow simulations. Table~\ref{tab:codes} summarizes state-of-the-art, architecture-agnostic software frameworks designed for high-order accurate simulations of three-dimensional, compressible flows on unstructured grids. SOD2D~\cite{gasparino2024sod2d} is a Fortran-based solver which uses OpenACC for multi-vendor support. It leverages a spectral element method (SEM) and offers an entropy viscosity stabilization approach to maintain robustness in the presence of discontinuities. To date, performance of SOD2D has been reported on up to 64 GPUs~\cite{eleftherakis2025poster}. In contrast, PyFR~\cite{witherden2025pyfr} is a Python-based framework centered on the flux reconstruction (FR) method in tandem with an entropy-based adaptive filtering approach for a-posteriori shock capturing~\cite{dzanic2023entropy}. By leveraging Mako templates~\cite{mako} to abstract GPU kernels in a vendor-agnostic way, PyFR achieves broad backend compatibility, including NVIDIA, AMD, Intel, and Apple Metal hardware and has demonstrated parallel performance up to 2048 GPUs.

While these frameworks offer robust solutions for turbulent, compressible flows, certain limitations currently restrict their broader applicability. A primary concern is the lack of proven performance at the largest computing scales. High-fidelity DNS and LES of turbulent flows at realistic Reynolds numbers often demand resolutions exceeding tens of billions of degrees of freedom (DOFs)~\cite{fujii2025scale,ceci2025grid}. Executing such simulations necessitates proven scalability not merely on thousands of GPUs, but on the order of tens of thousands.

Furthermore, the numerical fidelity and scalability of these solvers are significantly impacted by their respective numerical treatments of discontinuities. The entropy filtering approach in PyFR relies on a computationally demanding a posteriori sub-iteration process to adaptively determine the filter strength by enforcing positivity and local discrete minimum entropy principles~\cite{dzanic2023entropy}. This shares a fundamental limitation with artificial viscosity schemes: the inherent difficulty in retaining sub-cell resolution~\cite{Zeifang2021}. This often results in overly dissipative behavior that smears the solution, thereby degrading the high-order accuracy not just at the discontinuity, but in the smooth flow regions immediately surrounding the shock front. This is particularly crucial for more complex flow scenarios such as shock-turbulence interactions, where capturing the fine-scale amplification of eddies across a shock requires avoiding the overly dissipative behavior inherent to artificial viscosity or filter-based shock capturing approaches.

Finally, the specific architecture-agnostic strategies \\employed by these frameworks present hurdles for broader adoption and cross-platform transferability. OpenACC restricts \\SOD2D's effective portability to systems equipped with NVIDIA and AMD GPUs~\cite{herten2023gpumodels}, leaving a gap for other emerging hardware architectures. PyFR's reliance on a Python-centric stack is a distinct departure from the high-order CFD landscape, which is predominantly built upon the Fortran~\cite{fischer2007nek5000,devanna2023uranos,jansson2024neko,krais2021flexi,rubio2022horses3d} or C++~\cite{fischer2022nekrs} programming languages. Consequently, the specialized porting mechanisms developed for PyFR are not easily transferable to the majority of existing production codes, complicating the exchange of optimized kernels and limiting the wider impact of its portability breakthroughs.

To address the dual challenges of extreme scalability and hardware-agnostic portability without sacrificing high-order accuracy in the vicinity of discontinuities, this work introduces the final entry in Table~\ref{tab:codes}, \galaexi. This framework is built upon three core pillars:
\begin{itemize}
  \item \textbf{Broad and Transferable Portability:} \galaexi achieves portability by interfacing an existing Fortran code base to the C++ variants of the programming models CUDA~\cite{cudatoolkit} for NVIDIA GPUs and HIP~\cite{hiplatest} for AMD GPUs, providing a template for porting other Fortran-based computational software. This approach ensures immediate performance on NVIDIA and AMD hardware while remaining extensible to other vendors and hardware.
  \item \textbf{Scalability:} Leveraging the excellent parallelization properties of the discontinuous Galerkin spectral element \\method (DGSEM) --- which has demonstrated scalability to over \num{500000} CPU cores~\cite{blind2023towards} --- this work extends that efficiency to the GPU domain. This targets the deployment on tens of thousands of accelerators required for next-generation DNS and LES, moving beyond the proof-of-concept performance demonstrated in initial GPU porting attempts~\cite{kurz2024galaexi}.
  \item \textbf{High-Fidelity:} Unlike artificial viscosity or filter-based shock capturing approaches, \galaexi utilizes a more accurate shock capturing based on the switching of troubled DG cells to an equidistant finite volume (FV) sub-cell operator~\cite{sonntag2014shock}. This allows for robust handling of discontinuities and, combined with the DGSEM, strictly preserves high-order accuracy and sub-element resolution in the surrounding turbulent flow.
\end{itemize}

By unifying these three aspects, \galaexi provides a high-performance pathway for the high-fidelity simulation of complex, compressible flows at previously inaccessible scales. \galaexi is derived from the open-source solver \flexi~\cite{krais2021flexi}, a mature codebase thoroughly validated across a diverse range of fluid dynamics applications~\cite{beck2014high,Keim2026,Mossier2026}. From this foundation, \galaexi inherits a robust suite of high-order numerical algorithms. This work focuses its core advancements on elevating the capabilities of those algorithms to achieve performance and architecture-agnostic portability for complex, compressible flow simulations.

The discussions that follow are organized thus. The next section provides a review of architecture-agnostic GPU porting approaches for Fortran-based computational physics software. Section~\ref{sec:implementation} then describes the governing equations and numerics scheme of \galaexi before elaborating on the implemented approach for architecture-agnostic computing. The validation of the implementation is presented in Section~\ref{sec:validation}. The performance of \galaexi is highlighted in Section~\ref{sec:performance}, with aspects such as the scaling performance, time- and energy-to-solution discussed. Finally, in Section~\ref{sec:application}, \galaexi is used to compute a wall-resolved LES (WRLES) of two transonic airfoil configurations to demonstrate its performance for representative use cases. Section~\ref{sec:conclusion} then provides a summary and conclusion.

\section{Architecture-agnostic Acceleration Approaches for Fortran Software}
\label{sec:multibackend}

\begin{table*}[t]
\begin{minipage}{\textwidth}
  \centering
  \caption{
    Selection of available programming models for the architecture-agnostic GPU porting of Fortran-based scientific software
  }
  \label{tab:approaches}
  \begin{tabular}{|p{5cm}||p{2cm}|p{4.5cm}|p{2.5cm}|}
    \hline
    \thead{Programming\\Model} & \thead{Native Fortran\\Support} & \thead{Supported Vendors /\\Architectures} & \thead{Level of\\Abstraction} \\
    \hline
    \hline
    \makecell*[l]{OpenMP~\cite{openmp}} & YES & CPU, NVIDIA, AMD, INTEL & Directives \\
    \hline
    \makecell*[l]{OpenACC~\cite{openacc}} & YES & CPU, NVIDIA, AMD & Directives \\
    \hline
    \makecell*[l]{Kokkos~\cite{trott2022kokkos}} & NO & CPU, NVIDIA, AMD, INTEL & Framework \\
    \hline
    \makecell*[l]{RAJA~\cite{raja}} & NO & CPU, NVIDIA, AMD, INTEL & Framework \\
    \hline
    \makecell*[l]{ALPAKA~\cite{zenker2016alpaka}} & NO & CPU, NVIDIA, AMD & Framework \\
    \hline
    \makecell*[l]{Vendor models\\(CUDA~\cite{cudatoolkit},HIP~\cite{hiplatest}, SYCL~\cite{khronossycl})} & Mixed\footnote{Each vendor's model natively supports that vendor's hardware. HIP also supports NVIDIA GPUs~\cite{hiplatest}, and SYCL supports all vendors~\cite{herten2023gpumodels}.} & Mixed\footnote{CUDA natively supports Fortran with CUDA Fortran~\cite{cudafortran}. HIP and SYCL only natively support C/C++.} & None \\
    \hline
  \end{tabular}
\end{minipage}
\end{table*}

While Fortran has not been a mainstream programming language for many decades, it is still highly prevalent in the computational sciences and HPC~\cite{fortranlang}. As was highlighted in Section~\ref{sec:introduction}, that includes many high-order CFD software frameworks. In the current age of heterogenous computing environments, the process of selecting a portable GPU porting strategy can be a difficult task for Fortran-based CFD codes. While there are many resources available to guide developers in making this impactful decision, there are no objective rules on which will be best for a given piece of software. Each software must individually select an approach based on their own requirements. The task is complicated by the fact there are rarely opportunities to reconsider the choice of approach once developers have committed to it. In Table~\ref{tab:approaches}, a selection of the possible portable programming models available to developers of Fortran scientific software is presented.

Attractive options in this category are the directive-based models OpenMP~\cite{openmp} and OpenACC~\cite{openacc}, which use code decorators called \emph{directives} to build structured blocks of code called \emph{constructs} that can execute on the CPU or GPU. While directives can be a non-intrusive way to port Fortran codes to GPUs, algorithms may still demand extensive code refactors to account for the multi-threaded nature of GPUs~\cite{bak2022openmp}. Both OpenMP and OpenACC are Fortran-native, foregoing any need to rewrite portions of the code base in another programming language to support certain hardware or vendors. OpenMP supports offloading to GPUs from NVIDIA, AMD and Intel while OpenACC supports GPU hardware from NVIDIA and AMD~\cite{herten2023gpumodels}. The main drawbacks of directive models is their lack of fine-grain control over performance and their heavy reliance on compiler support for performance and portability. In a review of OpenMP porting efforts~\cite{bak2022openmp}, codes from many scientific disciplines found performant results, though some struggled with OpenMP's feature support across compilers and control over GPU memory.

Looking specifically at CFD software packages, several have used directive models to port their codes. SOD2D~\cite{gasparino2024sod2d} uses OpenACC and can compute on CPU-based systems and can also offload to NVIDIA and AMD GPUs. STREAmS-2~\cite{bernardini2021streams,salvadore2024streams} is a high-order finite difference (FD) solver which supports Intel, AMD and NVIDIA GPUs while retaining CPU support and has been demonstrated on simulations featuring more than 2000 GPUs. The developers of STREAmS-2 used a mixture of CUDA Fortran~\cite{cudafortran}, OpenMP and a bespoke source-to-source (S2S) translation tool to achieve this. Another solver is URANOS~\cite{devanna2023uranos,devanna2024uranos}, which supports computations on CPUs and GPU hardware from NVIDIA and AMD using OpenACC.

Another option to provide maximum portability are a class of methods called \emph{abstraction frameworks}. Examples of such frameworks include Kokkos~\cite{trott2022kokkos}, RAJA~\cite{raja} and ALPAKA~\cite{zenker2016alpaka}. All of these frameworks offer some level of vendor-agnostic abstraction, allowing GPU code to be written generally then the individual architectures are handled behind the scenes by the framework. The biggest drawback of these libraries, however, is their either cursory or non-existent Fortran support.

The final option is using the \emph{vendor models} offered by each GPU hardware manufacturer. Examples of these models include CUDA~\cite{cudatoolkit} for NVIDIA GPUs, HIP~\cite{hiplatest} for AMD and SYCL~\cite{khronossycl} for Intel. HIP also boasts support for NVIDIA GPUs~\cite{hiplatest} and use of SYCL for both AMD and NVIDIA GPUs is possible through the AdaptiveCpp project~\cite{solanki2025adaptivecpp}. NVIDIA's CUDA offers full support for GPU kernel code in Fortran~\cite{cudafortran}. The limitation here is that AMD and Intel do not offer this same support, with their models only available for use in C or C++ software~\cite{herten2023gpumodels}. From a performance perspective, vendor models allow developers to interact directly with the accelerator hardware with little to no abstraction, providing the highest degree of control. When compared in the context of a high-order DG code, OpenMP and Kokkos were slower than CUDA on NVIDIA GPUs~\cite{dai2024dgcomp}. The authors of \citet{malaya2023exascale} concluded that the largest performance benefits were seen when scientific codes used vendor models for porting.

If either the abstraction frameworks or vendor models are chosen, all compute methods would have to be rewritten in C/C++ in order to support all GPU vendors. Both NEKO~\cite{jansson2024neko} and FUN3D~\cite{nastac2021fun3dgpu} are examples of Fortran CFD codes that leverage vendor models through Fortran-to-C interfaces in a performant and portable way. NEKO is a GPU-accelerated version of the well-known CFD software Nek5000~\cite{fischer2007nek5000}. It utilizes a SEM, but is limited to the incompressible regime on structured grids. NEKO leverages modern, object-oriented Fortran to create a class hierarchy where backends for multiple vendors can be implemented as instances of solver classes. Through that approach, NEKO features support for CPUs and GPUs from NVIDIA, AMD and Intel and has been shown to scale up to 512 GPUs~\cite{jansson2025nekoperf}. The finite volume software package FUN3D~\cite{fun3d14manual} uses interfaces to the C++ vendor models CUDA~\cite{cudatoolkit}, HIP~\cite{hiplatest} and SYCL~\cite{khronossycl}, which are contained in an external library called FLUDA. With this approach, FUN3D has been scaled to thousands of GPUs from multiple vendors~\cite{nastac2021fun3dgpu,nastac2023multiarch}.

After weighing the benefits and considering the \\above-enumerated examples of each programming model, the approach using Fortran-C interfaces to the C++ vendor models was chosen for \galaexi. This is due to the available fine-grain control of performance, success of the approach in similar software packages and a proven record of scaling on the largest supercomputers in the world.

\section{Implementation}
\label{sec:implementation}
In the following section, brief coverage of the DGSEM \\method is provided. That is followed by a detailed description of \galaexi's strategy for interfacing an existing Fortran implementation of the DGSEM to the C++ vendor models including coverage on memory management and parallelization strategies. We introduce these numerical and implementation aspects to provide the necessary context for the performance benchmarks and physical validations that follow.

\subsection{Governing Equations and Numerical Method}
\label{subsec:numerics}

A compressible, viscous fluid governed by the \\Navier--Stokes--Fourier equations is considered, given as:

\begin{align}
  &\lr{\cons}_t + \nabla \cdot \mathbf{F}(\cons,\nabla \cons) = \mathbf{0}, \\
  &\mathbf{F} = \left[ \rho \mathbf{u}, \rho \mathbf{u} \otimes \mathbf{u} + p \mathbf{I} - \boldsymbol{\tau}, (\rho e + p)
  \mathbf{u} - \boldsymbol{\tau} \cdot \mathbf{u} - \lambda \nabla T \right]^\top,\nonumber
\end{align}

where $\cons=\left[ \rho, \rho \mathbf{u}, \rho e \right]^\top$, is the vector of conserved variables, consisting of the density $\rho$, the velocity vector $\mathbf{u}$, and the total energy $e$ per unit mass. The unit tensor is denoted by $\mathbf{I}$, $T$ is the temperature, $\lambda$ the thermal conductivity, $p$ the pressure, $\mathbf{F}$ the physical flux, and the viscous stress tensor $\boldsymbol{\tau}$ for a Newtonian fluid. The heat flux is modeled in accordance with Fourier's law. The equations are closed by the caloric and thermal equations of state of a perfect gas, given as $p = \rho (1-\gamma) \epsilon$ and $p=\rho R T$ with the specific gas constant $R$ of ambient air and the internal energy $\epsilon$ per unit mass.

Following the method of lines approach, the conservation equations are discretized in space by the DGSEM. The computational domain $\smash{\Omega \subseteq \mathbb{R}^3}$ is discretized by non-overlapping, (non-) conforming hexahedral elements with six possibly curvilinear element faces. Subsequently, the considered governing equations are mapped onto the reference space $E=[-1,1]^3$ via the mapping $\boldsymbol{\chi}: \boldsymbol{\xi} \in E \mapsto \mathbf{x} \in \Omega$. The Jacobian matrix of this mapping is $\mathbf{J} = \lr{\nabla_{\xi} \boldsymbol{\chi}} \in \mathbb{R}^{3\times 3}$ with $\mathcal{J}=\det\mathbf{J}$ as the corresponding determinant. The element-local solution $\cons$ is approximated by a polynomial representation using the tensor-product of one-dimensional nodal Lagrange basis functions $\ell$ of degree $N$, The weak form is retrieved by a discrete $L_2$ projection of the resulting equations onto the test space composed of polynomials $\testfunc(\boldsymbol{\xi})$ up to degree $N$, chosen similarly to the basis functions (Galerkin approach) and followed by an application of Gauss' theorem, yielding

\begin{align}
  \label{eq:dgsem_weak}
  \int_E \mathcal{J} \cons_t \testfunc(\boldsymbol{\xi}) \,d\boldsymbol{\xi}
  +& \int_{\partial E} (\fphysref\cdot\hat{\mathbf{n}})^\ast \testfunc(\boldsymbol{\xi}) \,d\mathbf{S} \\
  -& \int_E \fphysref(\cons, \nabla \cons) \cdot \nabla_\xi \testfunc(\boldsymbol{\xi}) \,d\boldsymbol{\xi} = 0, \nonumber
\end{align}

where $\hat{\mathbf{n}}$ is the unit normal vector in $E$, and $\fphysref$ are the contravariant fluxes. Neighboring elements are weakly coupled via the numerical flux $(\fphysref\cdot\hat{\mathbf{n}})^\ast$ normal to the element faces, here approximated by Roe's numerical flux with the entropy fix by~\cite{Harten1983b} and the viscous fluxes are simply averaged. The collocation property is exploited for the numerical integration of~\cref{eq:dgsem_weak} by the use of $(N+1)^3$ Legendre--Gauss--Lobatto or Legendre--Gauss quadrature points as interpolation nodes for $\cons$.

To calculate the viscous fluxes, the Navier--Stokes--Fourier equations require the gradients of the primitive variables, $\nabla \mathbf{q}^{\text{prim}}$. In this work, the BR1 lifting scheme \cite{Bassi1997} is applied to approximate these gradients. This approach involves solving an auxiliary set of equations; after applying the DGSEM, the resulting semi-discrete form is written as

\begin{align}
  \label{eq:dgsem_lifted}
  \int_E \mathcal{J} \mathbf{g} \testfunc(\boldsymbol{\xi}) \,d\boldsymbol{\xi} ~
  -& \int_{\partial E} \avg{\cons^{\text{prim}}\cdot\hat{\mathbf{n}}} \testfunc(\boldsymbol{\xi}) \,d\mathbf{S} \\
  +& \int_E \cons^{\text{prim}} \cdot \nabla_\xi \testfunc(\boldsymbol{\xi}) \,d\boldsymbol{\xi} = 0, \nonumber
\end{align}

with the lifted gradients $\mathbf{g} = \nabla_x \cons^{\text{prim}}$ and the averaging operator at the element faces $\avg{\cdot}_{ij} = [(\cdot)_i+(\cdot)_j]/2$.

In this work, the entropy-stable DGSEM with the entropy conservative fluxes of~\cite{Chandrashekar2013} is utilized to alleviate numerical stability issues and mitigate aliasing errors due to the approximation of the nonlinear (convective) flux. The shock capturing procedure is based on a 2nd-order accurate finite volume (FV) sub-cell scheme with $(N+1)^3$ integral means per DG element~\cite{Sonntag2017a} using a modal indicator~\cite{persson2006Indicator} for the detection of troubled DG cells. The solution is advanced in time by an explicit low-storage Runge--Kutta fourth-order accurate scheme with five stages (RK4-5)~\cite{Carpenter1994} or 14 stages (RK4-14)~\cite{Niegemann2012}. If not stated otherwise, in this work a $\mathrm{CFL}$ (Courant, Friedrichs and Lewy) number of $\mathrm{CFL}=0.9$ is set, the same applies to the viscous time step restriction. The reader is referred to~\cite{Sonntag2017a, krais2021flexi, Schwarz2025} for further details on the entropy-stable DGSEM, the FV sub-cell based shock capturing procedure, and applications.

\subsection{Porting Strategy}
\label{subsec:porting}

\galaexi inherits the Fortran core of its code base from the CPU-only DGSEM framework \flexi~\cite{krais2021flexi}. To support multiple vendors of GPU hardware, the Fortran code interfaces to compute kernels written in the C++ variants of the vendor models. To conquer the limitations of Fortran-to-C interoperability as well as the limitations of the boundary between CPU and GPU, the programming language barrier is placed at the hardware barrier. This results in a software design in \galaexi where all Fortran code executes on the CPU and all C/C++ code either executes on the GPU or mutates the GPU's state through kernel launches or vendor model API calls.

Owing to the similarities between CUDA C++ and HIP C++, the same kernel code can be reused for both CUDA and HIP, with the small differences between the two vendor models hidden behind a thin abstraction layer using pre-processor macros. New methods named \emph{entry points} branch between the different backends of a compute kernel at runtime. An example of such an entry point method is found in Listing~\ref{lst:entry_point}. Beyond their primary purpose of splitting between different hardware backends, the entry point methods can also perform operations that are shared between CPU and GPU backends.

\begin{lstlisting}[float,language=Fortran, caption=Fortran entry-point method to split between architecture backends in \galaexi. Note the use of a pre-processor flag to decide which backend will be invoked at compile time., label=lst:entry_point, linewidth=\columnwidth]
SUBROUTINE EntryPoint(nTotal,VecOut,d_Out,Const)
USE ISO_C_BINDING, ONLY: C_PTR
IMPLICIT NONE

! INPUT/OUTPUT VARIABLES
!< memory size length
INTEGER,INTENT(IN)      :: nTotal
!< host memory
REAL(KIND=8),INTENT(INOUT) :: VecOut(nTotal)
!< device memory
TYPE(C_PTR),INTENT(IN)  :: d_Out
!< constant
REAL(KIND=8),INTENT(IN) :: Const

#if (USE_ACCEL == ACCEL_OFF)
    CALL HostBackend(nTotal,VecOut,Const)
#else
    CALL DeviceBackend(nTotal,d_Out,Const)
#endif

END SUBROUTINE EntryPoint
\end{lstlisting}

For computations on CPUs, the original Fortran compute method from \flexi is called. When the GPU backends are active, the call to {\small\ttfamily DeviceBackend} in Listing~\ref{lst:entry_point} routes through a standard Fortran-C interface~\cite{reid2007fortran} and into a method called the \emph{kernel launcher}, an example of which is shown in Listing~\ref{lst:launcher}. The {\small\ttfamily INVOKE\_KERNEL} macro in Listing~\ref{lst:launcher} hides the vendor-specific kernel launch syntax.

\begin{lstlisting}[float,language=C++, caption=Example of a kernel launcher method in \galaexi. The {\small\ttfamily INVOKE\_KERNEL} macro hides the vendor-specific kernel launch syntax., label=lst:launcher, linewidth=\columnwidth]
#if (USE_ACCEL == ACCEL_CUDA)
#define INVOKE_KERNEL(fun,blcks,thrds,shrd,strm,...)\
    fun<<<blcks,thrds,shrd,strm>>>(__VA_ARGS__)
#elif (USE_ACCEL == ACCEL_HIP)
// Define AMD kernel launch syntax
#endif /* USE_ACCEL */

void DeviceBackend(int nTotal,void* d_VecOut,double Const)
{
    INVOKE_KERNEL(KernelFunc, nTotal/256+1,256,0,0,
        nTotal, (double*)d_VecOut, Const
    );
}
\end{lstlisting}

\subsubsection{Memory Management}
\label{subsubsec:memory}

A major design challenge with the chosen approach was how to manage GPU memory. With all initialization routines remaining in Fortran, the instructions to mutate GPU memory (i.e. allocate, free or copy to/from the CPU) must be issued from Fortran. While CUDA and HIP supply Fortran wrappers for their memory APIs, this would limit flexibility, as some models, such as SYCL, do not have Fortran wrappers. The approach seen in Listing~\ref{lst:mem_handling} is used, which keeps all backend-specific API calls in C++.

\vspace{\baselineskip}
\noindent
\begin{minipage}{\columnwidth}
\begin{lstlisting}[language=Fortran,linewidth=\columnwidth]
ALLOCATE( HostArr(arrSize) )
d_DevPtr = AllocateMemDevice(SIZE_DBLE,arrSize)
CALL CopyToDevice(d_DevPtr,HostArr,SIZE_DBLE,arrSize)
\end{lstlisting}
\begin{lstlisting}[language=C++, caption=Example of how GPU memory is allocated and managed in \galaexi. The top code snippet shows how the methods are called from Fortran. The bottom shows the API calls in C++ code. Pre-processor macros determine which API is used to allocate memory. This same design is used to expose all necessary functionality from vendor-specific C++ APIs in Fortran code., label=lst:mem_handling, linewidth=\columnwidth]
void* AllocateMemDevice(
    size_t typeSize_bytes, int arraySize
)
{
    void* d_arr;

#if (USE_ACCEL == ACCEL_CUDA)
    cudaMalloc(&d_arr, typeSize_bytes*arraySize);
#elif (USE_ACCEL == ACCEL_HIP)
    hipMalloc( &d_arr, typeSize_bytes*arraySize);
#endif /* USE_ACCEL */

    return d_arr;
}


void CopyToDevice(
    void* dVarPtr,void* hostMemory,
    size_t typeSize_bytes,int arraySize
)
{
    // ... CUDA/HIP memory copy calls
}
\end{lstlisting}
\end{minipage}

Fortran code calls the wrapper API for GPU memory handling. The call passes to a C++ function which uses the backend appropriate memory allocator (e.g. \code{cudaMalloc}) to allocate GPU memory. A pointer to this newly allocated GPU memory is returned to the Fortran side and stored in a Fortran variable. There it can be passed through functions in Fortran code and handed off to the C++ side later to perform operations on the data, such as memory copies and kernel launches. MPI communication routines and other vendor model API calls for device synchronization, pinning, etc. are abstracted in the same fashion.

As directly mapping runtime allocated, assumed-shape Fortran arrays to multi-dimensional C arrays is not possible~\cite{reid2007fortran}, all CPU-allocated Fortran arrays are mapped to flattened, one-dimensional arrays in GPU memory. To avoid reordering operations when copying data between the hardware and programming languages, data retains the original CPU allocated, Fortran column major ordering in GPU kernels. A small set of indexing functions is used in kernel code to find the offset to a particular thread's data in the flattened arrays.

\subsubsection{Thread Level Parallelism}
\label{subsubsec:threading}
The thread parallelization strategy of GPU kernels in \galaexi, as is common in GPU-accelerated DG codes~\cite{klockner2009dg,chan2016dg,karakus2019dg,kurz2024galaexi}, assumes a single GPU thread operates on a single degree of freedom (DOF). For the core operations of the DGSEM, this approach is appropriate, as many of those operations are pointwise, or \emph{DOF-local}. This means individual DOFs can complete operations independently of all others without the need to handle data dependencies between GPU threads.

\begin{algorithm}[tb]
  \caption{Face-local algorithm for projecting solution from big to small faces for two-to-one non-conforming element (mortar) interfaces on the CPU}
  \label{alg:fillmortar_cpu}
  \begin{algorithmic}[1]
    \Function{FillMortarFaces}{$N,n_{Mortars},\ppvec{U},\ppvec{I_L},\ppvec{I_R},\ppvec{Mort_{map}}$}
      \For{$n \gets 1\text{ to }n_{Mortars}$} \Comment{loop over mortar faces}
        \State ! Perform mapping
        \For{$p,q \gets 0\text{ to }N$} \Comment{loop over face DOFs}
          \For{$r \gets 0\text{ to }N$}
            \State ! Interpolate (non-DOF local operation)
            \State $\ppvec{U}^{tmp}_{pq,1} \gets \ppvec{U}^{tmp}_{pq,1} + \ppvec{U}_{pr,n}\ppvec{I}_{L_{rq}}$
            \State $\ppvec{U}^{tmp}_{pq,2} \gets \ppvec{U}^{tmp}_{pq,2} + \ppvec{U}_{pr,n}\ppvec{I}_{R_{rq}}$
          \EndFor
        \EndFor

        \State ! Fill solution big to small
        \For{$p,q \gets 0\text{ to }N$}\Comment{loop over face DOFs}
          \For{$i \gets 1\text{ to }2$}\Comment{two small sides per large}
            \State $j \gets \ppvec{Mort}_{{map}_i}$
            \State $\ppvec{U}_{pq,j} \gets \ppvec{U}^{tmp}_{pq,i}$
          \EndFor
        \EndFor
      \EndFor
      \State \Return {$\ppvec{U}$}
    \EndFunction
  \end{algorithmic}
\end{algorithm}

\begin{table}[h]
  \centering
  \caption{
    Comparison of wall times for two configurations of the kernel used to fill fluxes at non-conforming mesh interfaces in \galaexi. The times were taken on a single NVIDIA GeForce 4070 Ti Super GPU for 32 total non-conforming faces with a polynomial order of $N=7$. Times were collected using NVIDIA's NSight Compute \cite{nsightcompute} profiling software.
  }
  \label{tab:threading_comp}
  \begin{tabular}{|l||c|c|}
    \hline
    \thead{Approach} & \thead{Operation\\ Wall Time [ms]} & \thead{Speedup} \\
    \hline
    Face-local kernel & \num{1.888} & -- \\
    \hline
    Multiple DOF-local kernels & \num{0.06392} & 29.54x \\
    \hline
  \end{tabular}
\end{table}

However, some operations in \galaexi are \emph{not} DOF-local. It was found that breaking such non-DOF-local operations into multiple, DOF-local kernels was significantly more performant that having one large non-DOF-local kernel. To emphasize, the procedure for filling non-conforming element interfaces is presented as an example. \galaexi allows meshes to contain non-conforming element faces (i.e. hanging nodes), using the mortar technique of \citet{kopriva1996conservative}. Algorithm~\ref{alg:fillmortar_cpu} presents the \emph{face-local} process used on the CPU to map the solution across such interfaces. The non-DOF local aspect of the algorithm is the filling of $\ppvec{U}$ for a given DOF $p,q$ depends on result of the first tensor product, $\ppvec{U}^{tmp}$, for other DOFs in the face. Breaking Algorithm~\ref{alg:fillmortar_cpu} into two DOF-local operations involves each loop over the DOFs in a face becoming its own operation. Every face must finish calculating the intermediate solution $\ppvec{U}^{tmp}$ and store it in global memory before the filling of $\ppvec{U}$ can proceed.

Both face-local and DOF-local versions of Algorithm~\ref{alg:fillmortar_cpu} were implemented as GPU kernels and tested. A comparison of the performance for the two approaches is presented in Table~\ref{tab:threading_comp}. The timings in the table were measured on a single NVIDIA GeForce 4070 Ti Super GPU using NVIDIA's NSight Compute~\cite{nsightcompute} profiling software. The computational mesh contained 32 non-conforming faces and the polynomial order was set to $N=7$, resulting in 64 DOFs per non-conforming interface and \num{2048} total DOFs on all non-conforming interfaces. Translating Algorithm~\ref{alg:fillmortar_cpu} directly to a GPU kernel which assigns each thread a single \emph{face} (i.e. one thread operates on 64 DOFs), the operation completes in \num{1.888}ms. Using the split version of Algorithm~\ref{alg:fillmortar_cpu} with two DOF-local GPU kernels results in a total runtime of 0.06392ms, a 29.54x speedup over the single face-local kernel. That runtime is for both DOF-local kernels combined, including launch latencies and tail effect (i.e. the time required to complete the last partial wave of computations in a kernel where GPU utilization is low). In general, any algorithm in \galaexi that requires an intermediate value be calculated for multiple DOFs and then accessed in a non-DOF-local fashion by subsequent steps uses this kernel splitting approach. Some examples include the calculation of the modal troubled cell indicator and the switching operation of troubled DG cells to FV sub-cells.

\subsubsection{Intra-GPU Parallelism}
\label{subsubsec:intragpu}

\begin{figure*}
  \centering
  \includegraphics[width=0.9\textwidth]{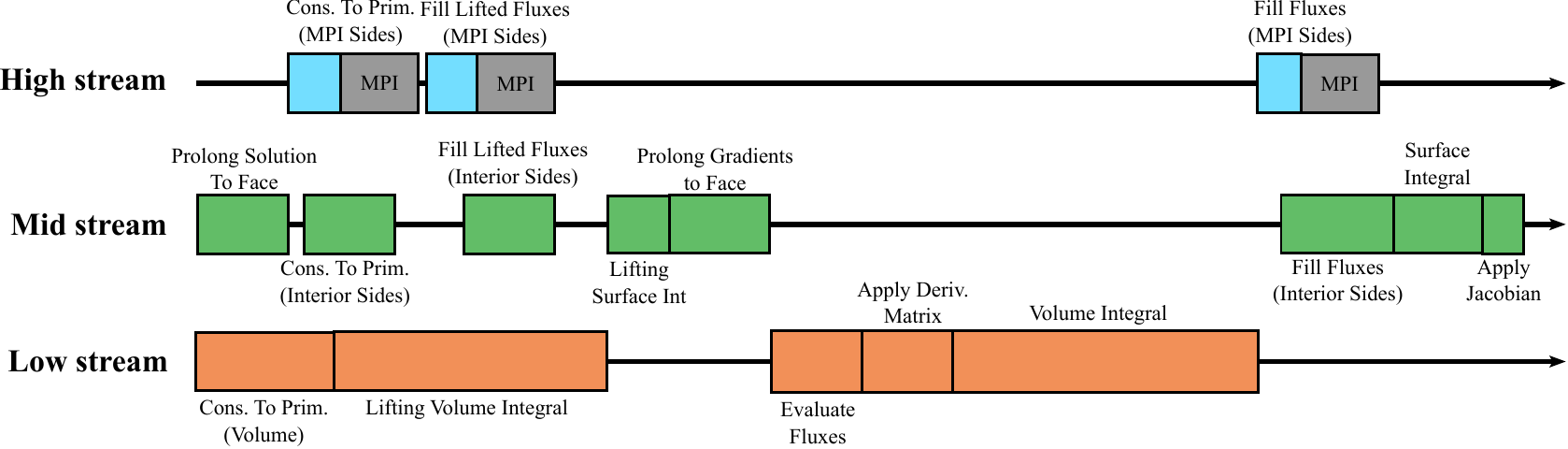}
  \caption{
    Evaluation timeline of the entropy-stable DGSEM from~\cref{eq:dgsem_lifted} including the lifting operator with device streams in \galaexi. Computation proceeds from left to right. Operations on data needed for MPI communication are assigned to the highest priority stream. Large volumetric operations, such as the volume integration, are given lowest priority. All other operations are given middle priority.
  }
  \label{fig:dgsem_timeline}
\end{figure*}

To maximize occupancy, \emph{device streams} are used on both AMD and NVIDIA GPUs. Detail of the DGSEM operation's distribution across the device streams is shown in Figure~\ref{fig:dgsem_timeline}. Summarized briefly, some operations in the DGSEM are input data-independent and can be performed simultaneously on two different device streams. An example is the conversion of the conservative variables to primitive variables (Cons. To Prim. (Volume) in Figure~\ref{fig:dgsem_timeline}) in the volume and the prolongation of the conservative variables to the element faces (Prolong Solution To Face in Figure~\ref{fig:dgsem_timeline}). \galaexi uses three device streams with three levels of increasing priority. Kernels that operate on element faces located on the boundary between MPI processes are the highest priority, to ensure data for MPI communication is ready as early as possible. The lowest priority is for volume operations within elements, which are independent of neighboring elements. The volume operations can compute in the background while the operations on element faces occur on the higher priority streams.

A challenge with using device streams is that there is no guarantee that upstream data dependencies are satisfied before a kernel begins. To enforce data dependencies between kernels, \emph{device events} are used. Device events are a CUDA/HIP construct that allows markers to be attached to a kernel and used downstream in the computation to query the execution status of the kernel~\cite{cudatoolkit,hiplatest}. In \galaexi, all kernels have an associated device event. Before launching a kernel, event synchronizations are used to confirm upstream kernels that operate on that kernel's input data are complete.

\subsubsection{Inter-GPU Parallelism}
\label{subsubsec:intergpu}

The final level of parallelism is that on distributed systems. The computational domain is divided into non-overlapping subdomains using a space-filling curve. Each subdomain is assigned to a single MPI process, which belongs to a single CPU core for CPU backends and a CPU core-GPU pair for device-accelerated backends. Non-blocking, point-to-point MPI communication is used to pass data between element faces that lie on subdomain boundaries. As with dependent kernels, device event synchronizations are used to check data for MPI communication is ready before the passing step begins. A representation of how MPI communication features in the runtime of \galaexi is included in Figure~\ref{fig:dgsem_timeline}. In \galaexi, device pinning, or the assignment of a GPU to a given MPI process/CPU core, is performed in-software using CUDA/HIP API calls.

\section{Validation}
\label{sec:validation}

In the following section, the implemented numerical methods are validated, starting with a generic manufactured solution to demonstrate
the convergence properties of the DGSEM. The discussion then progresses to the compressible Taylor--Green--Vortex to evaluate the scheme's robustness and accuracy in the context of shock-turbulence interactions.

\subsection{Convergence}
\label{subsec:convtest}

\begin{table*}[h]
  \centering
  \caption{
    \textit{h}-convergence of \galaexi for a manufactured solution with a polynomial order of $N=3$. Meshes were refined from a single element to $8^3$ elements.
  }
  \label{tab:h_conv}
  \begin{tabular}{|l||cc|cc|cc|}
    \hline
    \multirow{2}{*}{\thead{Grid}} & \multicolumn{2}{c|}{\thead{CPU}} & \multicolumn{2}{c|}{\thead{NVIDIA GH200}} & \multicolumn{2}{c|}{\thead{AMD MI250X}}\\
     & \thead{$L_2$ Error} &  \thead{EOC} & \thead{$L_2$ Error} & \thead{EOC} & \thead{$L_2$ Error} & \thead{EOC} \\
    \hline
    $1^3$ & \numThreePlaces{7.652821608e-2} & -- &
        \numThreePlaces{7.652821608e-2} & -- &
        \numThreePlaces{7.652821608e-2} & -- \\
    $2^3$ & \numThreePlaces{3.392616928e-3} & \numThreePlaces{4.495521325222666} &
        \numThreePlaces{3.392616928e-3} & \numThreePlaces{4.495521325222666} &
        \numThreePlaces{3.392616928e-3} & \numThreePlaces{4.495521325222666} \\
    $4^3$ & \numThreePlaces{3.030298652e-4} & \numThreePlaces{3.484866648541525} &
        \numThreePlaces{3.030298652e-4} & \numThreePlaces{3.484866648541525} &
        \numThreePlaces{3.030298652e-4} & \numThreePlaces{3.484866648541525} \\
    $8^3$ & \numThreePlaces{1.473344626e-5} & \numThreePlaces{4.362293154152285} &
        \numThreePlaces{1.473344626e-5} & \numThreePlaces{4.362293154152285} &
        \numThreePlaces{1.473344626e-5} & \numThreePlaces{4.362293154152285} \\
    \hline
  \end{tabular}
\end{table*}

To verify the implementation of the DGSEM in \galaexi, the spatial convergence of the scheme is rigorously tested using the method of manufactured solutions~\cite{roache2002code}. A periodic, rectilinear domain $\Omega = [-1,1]^3$ is initialized with a travelling, three-dimensional, sinusoidal solution of the form

\begin{align}
  \rho = 2+A\sin(2\pi(x+y+z-at)), \ \rho \mathbf{u} = \rho, \ \rho e = (\rho)^2,
\end{align}

where the amplitude and advection speed are chosen as $A=0.1$ and $a=1$, respectively. The solution is advanced in time to $t=1$. The timestep is chosen sufficiently small as to have no influence on the discretization error.

Table~\ref{tab:h_conv} presents the \textit{h}-convergence behavior of the $L_2$ error of the density on Gauss--Legendre quadrature nodes for a polynomial order of $N=3$. As the computational mesh is incrementally refined from a single element up to $8^3$ elements, the results confirm the theoretical order of convergence and the values for the $L_2$ error match to double precision across all three compared architectures.
For the $p$-convergence study, a mesh size of $4^3$ elements was held fixed and the polynomial order $N \in [4,8]$ gradually
increased. Again the values for the $L_2$ error of the density match to double precision for all architectures and demonstrate the expected exponential decay with increasing polynomial degree, cf. Table~\ref{tab:p_conv}.

\begin{table}[h]
  \centering
  \caption{
    \textit{p}-convergence of \galaexi using a manufactured solution on a $4^3$ element mesh.
  }
  \label{tab:p_conv}
  \begin{tabular}{|l||ccc|}
    \hline
    \multirow{2}{*}{\thead{N}} & \multicolumn{3}{c|}{\thead{$L_2$ Error}} \\
                               & \thead{CPU} & \thead{NVIDIA GH200} & \thead{AMD MI250X}\\
    \hline
    4 & \numThreePlaces{1.789152253e-05} & \numThreePlaces{1.789152253e-05} & \numThreePlaces{1.789152253e-05} \\
    5 & \numThreePlaces{3.62272299e-06}  & \numThreePlaces{3.62272299e-06 } & \numThreePlaces{3.62272299e-06 } \\
    6 & \numThreePlaces{1.045740633e-07} & \numThreePlaces{1.045740633e-07} & \numThreePlaces{1.045740633e-07} \\
    7 & \numThreePlaces{4.673515175e-08} & \numThreePlaces{4.673515182e-08} & \numThreePlaces{4.673515174e-08} \\
    8 & \numThreePlaces{7.472689016e-10} & \numThreePlaces{7.472689366e-10} & \numThreePlaces{7.472688652e-10} \\
    9 & \numThreePlaces{3.63433108e-10}  & \numThreePlaces{3.63433289e-10 } & \numThreePlaces{3.634332642e-10} \\
    \hline
  \end{tabular}
\end{table}

\subsection{Compressible Taylor-Green-Vortex}
\label{subsec:tgv}

The compressible Taylor--Green-Vortex~(TGV) at a Mach number of $\ppMa_0=1.25$ is a common benchmark for assessing the performance of numerical
schemes in capturing small scale turbulent structures while providing necessary dissipation to stabilize localized discontinuities~\cite{Lusher2021}.
This makes the compressible TGV particularly suited to examine the ability of \galaexi to simulate problems with complex shock-turbulence interactions.

Initial conditions for the velocity and pressure fields are given analytically as

\begin{align}
    \label{eq:tgv_ic}
    \ppvec{u}(\ppvec{x},0) &=
        \left(
        \begin{matrix}
        \phantom{-}U_0 \sin\left(\frac{x}{L}\right)\cos\left(\frac{y}{L}\right)\cos\left(\frac{z}{L}\right)\\
                    -U_0 \cos\left(\frac{x}{L}\right)\sin\left(\frac{y}{L}\right)\cos\left(\frac{z}{L}\right)\\
                    0
        \end{matrix}
        \right),\\
    p(\ppvec{x},0) &= p_0 + \frac{\rho_0 U_0^2}{16} \left(\cos\left(\tfrac{2x}{L}\right)+\cos\left(\tfrac{2y}{L}\right)\right)\left(2+\cos\left(\tfrac{2z}{L}\right)\right)\nonumber,
\end{align}

where $L=1$ is the characteristic length, $U_0=1$ is the magnitude of the initial velocity fluctuations and $\rho_0=1$ the reference density. The initial pressure $p_0$ is chosen to match a prescribed background Mach number $\ppMa_0=U_0\sqrt{\rho_0/(\gamma p_0)}$. The domain is a periodic cube with dimensions $\Omega\in\left[0,2{\pi}L\right]^3$.
To yield a complete set of initial conditions for a compressible flow field, the density is specified using the equation of state of an ideal gas $\rho(\ppvec{x},0) = \frac{p}{RT_0}$ with the temperature $T(\mathbf{x},t=0)=T_0$.
The desired Reynolds number, chosen as $\ppRe = \nicefrac{\rho_0 U_0 L}{\mu_0} = \num{1600}$, is obtained by adjusting the initial dynamic viscosity $\mu_0$.

The accuracy of the numerical scheme is assessed by two commonly used metrics, the integral kinetic energy,
\begin{equation}
  E_k = \frac{1}{2 \rho_{0} U_0^2\left|\Omega\right|} \int_\Omega \rho \, \ppvec{u} \cdot \ppvec{u} \,\mathrm{d}\Omega,
\end{equation}
and the solenoidal component of the viscous dissipation rate, defined in \citet{sarkar1991analysis} as
\begin{align}
  \varepsilon_S &= \frac{L^2}{ \ppRe U_0^2 \left|\Omega\right|} \int_\Omega \frac{\mu(T)}{\mu_0} \,\ppvec{\omega} \cdot \ppvec{\omega}\,\mathrm{d}\Omega,
\end{align}
where $\left|\Omega\right|$ is the total volume of the computational domain, $\mu(T)$ is the temperature-dependent dynamic viscosity of the fluid, computed using Sutherland's law~\cite{Sutherland1893} and $\omega$ is the vorticity vector.

\begin{figure}
  \centering
  \includegraphics[width=\columnwidth]{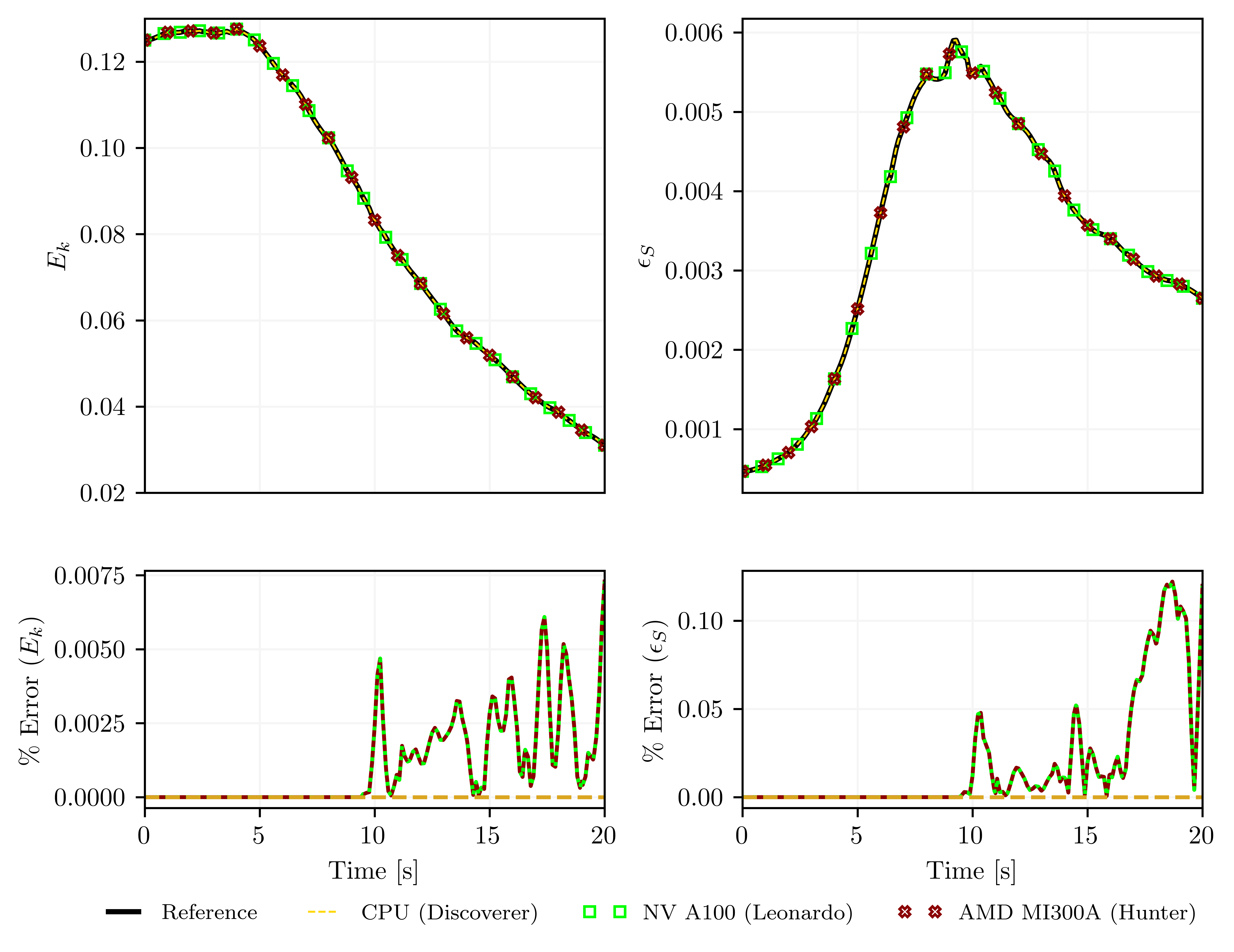}
  \caption{
    Comparison of the integral kinetic energy $E_k$ (top left) and solenoidal viscous dissipation rate $\varepsilon_S$ (top right) for the supersonic, compressible TGV. All simulations used a mesh with $16^3$ elements and a polynomial order of $N=3$. CPU computations for \flexi and \galaexi were performed on AMD EPYC CPUs. The results from \flexi are presented as the reference solution. The results for the GPU backends in \galaexi are shown for AMD MI300A APUs and NVIDIA A100 GPUs. The percent error in the solutions with \galaexi on the selected hardware is also presented for $E_k$ (bottom left) and $\varepsilon_S$ (bottom right)
  }
  \label{fig:comp_tgv}
\end{figure}

\begin{figure}
  \centering
  \includegraphics[width=0.9\columnwidth]{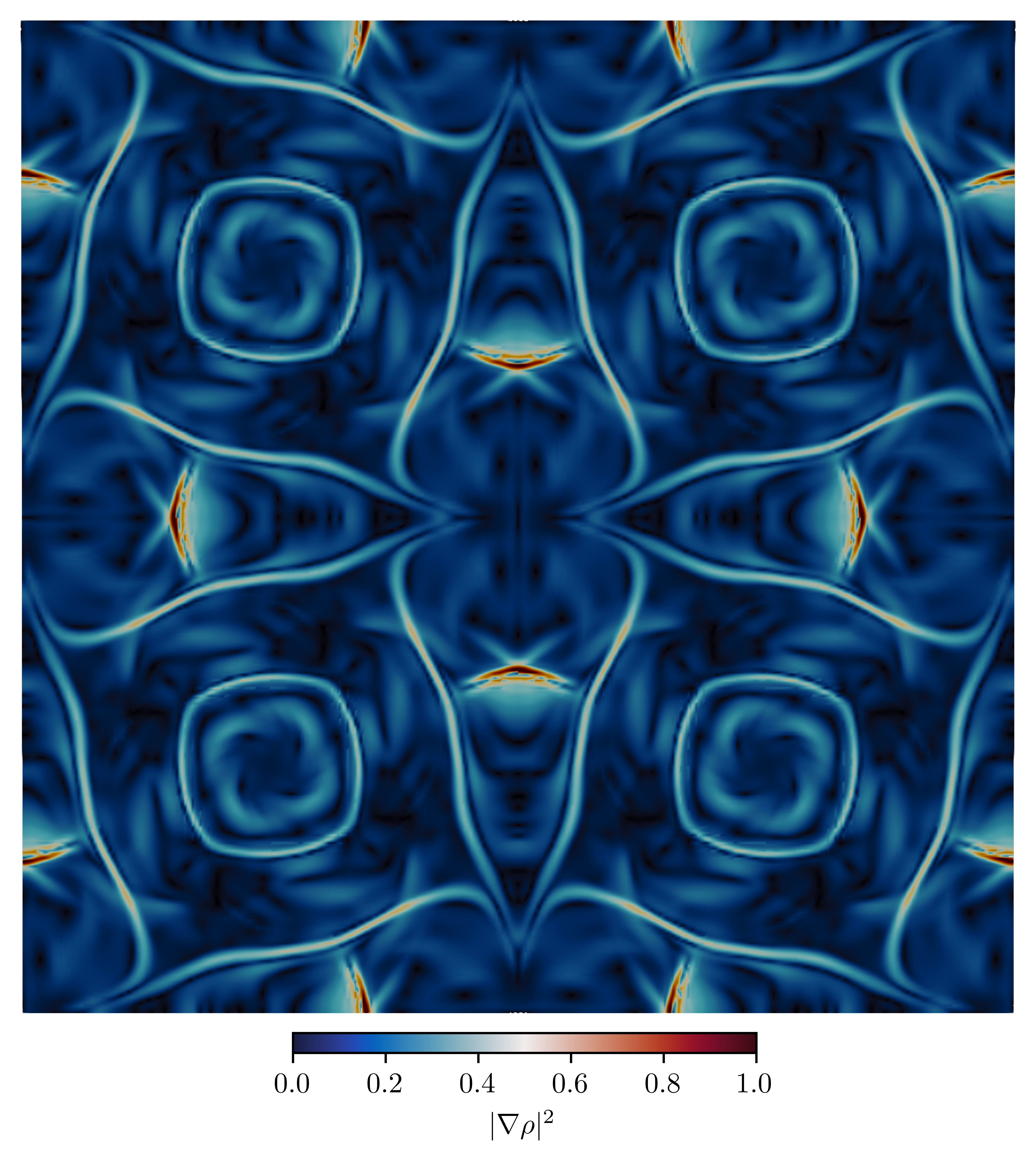}
  \caption{
    Center plane slice of a Schlieren image of the flow field for the compressible TGV case with $\ppMa_0=1.25$ and $\ppRe=\num{1600}$ at time $t=15$. Presented solution was computed with NVIDIA A100 GPUs on the Leonardo Booster supercomputer.
  }
  \label{fig:comp_tgv_vis}
\end{figure}

Computations were performed on a $16^3$ element mesh with a polynomial order of $N=3$. To suppress aliasing errors and ensure numerical stability in the presence of strong gradients, the entropy-stable DGSEM on Gauss--Legendre nodes coupled with a FV sub-cell based shock capturing approach was employed. CPU results were collected on the Discoverer supercomputer~\cite{discoverer} in Sophia, Bulgaria, with the solution from the CPU-only DGSEM framework \flexi serving as the reference. The GPU computations with \galaexi were performed with the AMD MI300A APUs on the Hunter supercomputer~\cite{hunter} at the High Performance Computing Center Stuttgart (HLRS) and the NVIDIA A100 GPUs on the Booster partition of the Leonardo supercomputer~\cite{leonardo} at CINECA in Bologna, Italy.

The results of the compressible TGV are given in Figure~\ref{fig:comp_tgv}. The bottom two plots of Figure~\ref{fig:comp_tgv} provide the percent error in $E_k$ and $\varepsilon_S$ of the solutions from \galaexi relative to the reference solution. The maximum percent error for the CPU backends of \galaexi is \numThreePlaces{1.9058301923800854e-11}\% and \numThreePlaces{7.899352609307615e-10}\% for $E_k$ and $\varepsilon_S$, respectively. For the GPU backends, the NVIDIA A100s and AMD MI300 produced comparable solutions, both with maximum percent errors of \numThreePlaces{0.0072870684229311864}\% for $E_k$ and \numThreePlaces{0.12217975596560209}\% for $\varepsilon_S$. A two-dimensional slice of the instantaneous solution at $t=15$ is given in Figure~\ref{fig:comp_tgv_vis}. The slice was taken at the center plane of the domain and shows a Schlieren image of the flow field using the result from Leonardo.

\section{Performance Evaluation}
\label{sec:performance}

\begin{table*}[t]
  \centering
  \caption{
    Hardware overview for the systems used for \galaexi development and testing. Due to the amount of systems, only the most basic information can be included. Note that the listed "CPU Node Hardware" corresponds to the CPU-only compute nodes. For some systems, the host CPU architecture on GPU-accelerated nodes differs from the CPU-only compute nodes. For those systems that do not have CPU-only compute nodes, the "CPU Node Hardware" is left blank. Detailed information on individual systems is available in the cited references.
  }
  \label{tab:systems}
  \begin{tabular}{|p{3cm}||p{1.5cm}|p{2.1cm}|p{2.5cm}|p{1.25cm}|p{3cm}|p{1.25cm}|}
    \hline
    \thead{System Name} & \thead{Computing\\Center} & \thead{Location} & \thead{CPU Node\\Hardware} & \thead{\# CPUs} & \thead{GPU Node\\Hardware} & \thead{\# GPUs} \\
    \hline
    \hline
    \makecell*[l]{Deucalion x86~\cite{deucalion}} & MACC & Guimarães, PT & AMD EPYC & \makecell{\num{64000}} & NVIDIA A100 & \makecell{132} \\
    \hline
    \makecell*[l]{Discoverer~\cite{discoverer}} & SofiaTech & Sofia, BG & AMD EPYC & \makecell{\num{144384}} & NVIDIA H200 & \makecell{32} \\
    \hline
    \makecell*[l]{Frontier~\cite{frontier}} & OLCF & Oak Ridge,\newline TN, USA & \makecell{--} & \makecell{--} & AMD MI250X & \makecell{78848} \\
    \hline
    \makecell*[l]{Hunter~\cite{hunter}} & HLRS & Stuttgart, DE & \makecell{--} & \makecell{--} & AMD MI300A & \makecell{\num{752}} \\
    \hline
    \makecell*[l]{JUPITER~\cite{jupiter}} & FZJ & J{\"u}lich, DE & \makecell{--} & \makecell{--} & NVIDIA GH200 & \makecell{\num{23536}} \\
    \hline
    \makecell*[l]{Karolina~\cite{karolina}} & IT4I & Ostrava, CZ & AMD EPYC & \makecell{\num{92160}} & NVIDIA A100 & \makecell{576} \\
    \hline
    \makecell*[l]{Leonardo~\cite{leonardo}} & CINECA & Bologna, IT & Intel Xeon & \makecell{\num{172032}} & NVIDIA A100 & \makecell{\num{13824}} \\
    \hline
    \makecell*[l]{LUMI~\cite{lumi}} & CSC & Kajaani, FI & AMD EPYC & \makecell{\num{262144}} & AMD MI250X & \makecell{\num{23824}}\\
    \hline
    \makecell*[l]{Marenostrum5~\cite{marenostrum}}& BSC & Barcelona, ES & Intel Xeon & \makecell{\num{726880}} & NVIDIA H100 & \makecell{\num{4480}} \\
    \hline
    \makecell*[l]{MeluXina~\cite{meluxina}} & LuxProvide & Bertrange, LU & AMD EPYC & \makecell{\num{73216}} & NVIDIA A100 & \makecell{800} \\
    \hline
    \makecell*[l]{Vega~\cite{vega}} & IZUM & Maribor, SI & AMD EPYC & \makecell{\num{98304}} & NVIDIA A100 & \makecell{240} \\
    \hline
  \end{tabular}
\end{table*}

To demonstrate the portability of \galaexi, it was deployed on eleven HPC systems. The hardware and configuration of each of these systems is summarized in Table~\ref{tab:systems}. Throughout the remaining sections of this publication these systems will be referred to by the \emph{System Name} listed in Table~\ref{tab:systems}.

Please note that the AMD MI250X is a multi-chip module (MCM) and a single MI250X contains two GPU dies, called graphics compute devices (GCDs)~\cite{mi250x_datasheet, lumi}. Therefore, in Table~\ref{tab:systems} and throughout the rest of this work, when the number of MI250X GPUs is mentioned, this is the number of total GCDs, which is always twice the number of AMD MI250X modules.

\subsection{Performance Metrics}
\label{subsec:perf_metrics}
The primary performance metric for comparing backends of \galaexi is the \emph{time-to-solution}, which the is total wall time required for a computation in \galaexi to complete, including the time required for operations such as data initialization and file output. In some cases, another metric known as the \emph{performance index~(PID)} is used. The PID represents the time to integrate a single DOF through single RK stage on a single CPU core or single GPU and is defined as
\begin{equation}
  \text{PID} = \frac{\text{Walltime}\; \times\; \#\text{Ranks}}{\#\text{RK-stages}\; \times\; \#\text{DOF}}.
\end{equation}

When energy consumption is discussed, total power draw of a simulation will be given in megajoules~(MJ). Where available, the \code{sacct} tool from SLURM~\cite{slurm} was used to report energy usage. The exception to this was on Marenostrum 5, where the EAR~\cite{ear} framework was used.

\subsection{Scaling Performance on GPU Systems}
\label{subsec:gpu_perf}

\begin{figure*}
  \centering
  \includegraphics[width=0.9\textwidth]{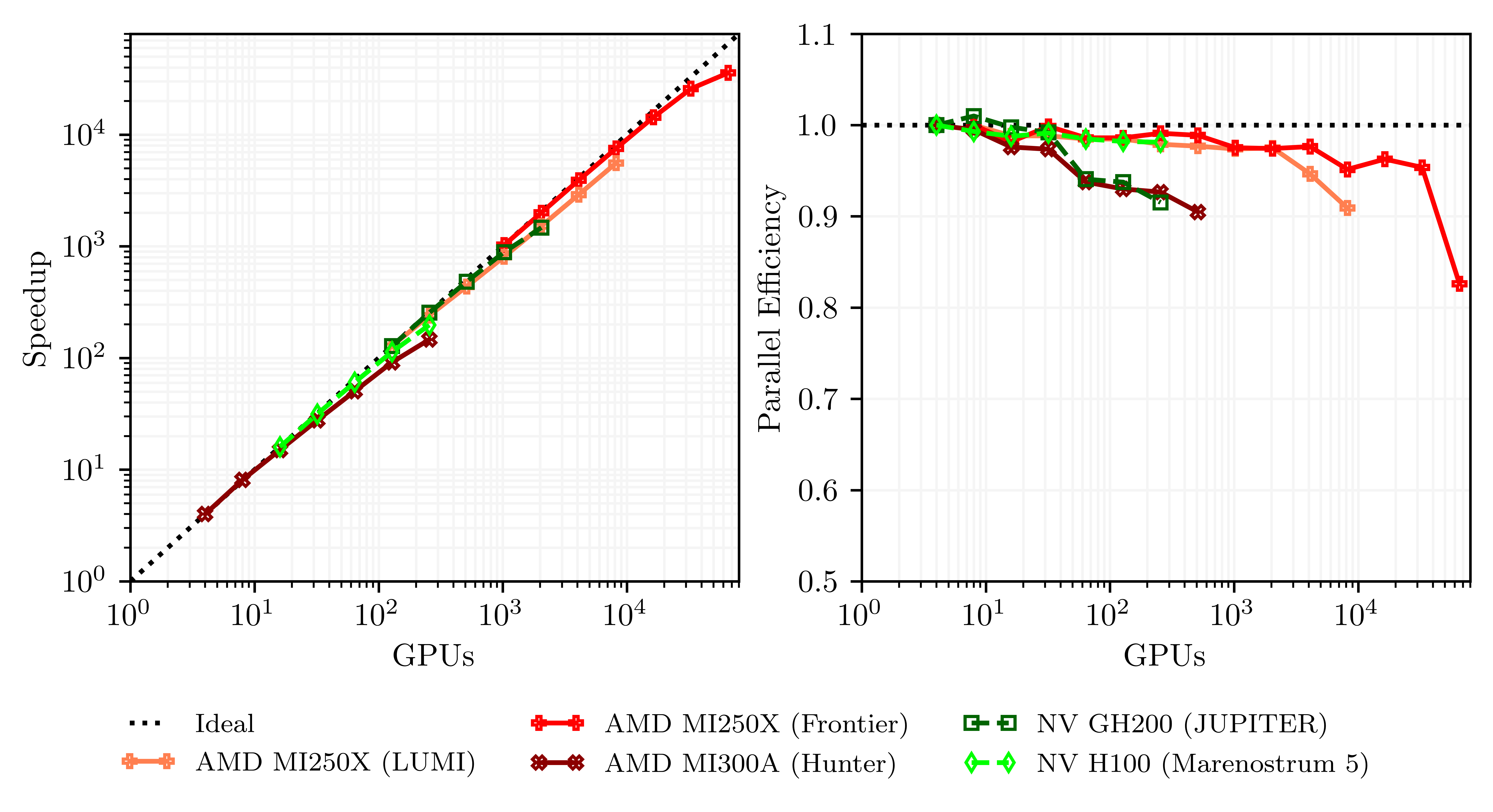}
  \caption{
    Strong (left) and weak (right) scaling performance of \galaexi on GPU-accelerated HPC systems.
  }
  \label{fig:gpu_scaling}
\end{figure*}

The parallel performance of \galaexi on GPU-accelerated systems was investigated using strong and weak scaling experiments. For both weak and strong scaling, a simulation of a freestream flow is advanced ten timesteps on a Cartesian mesh. The tests are run with the split-flux formulation of the DGSEM on Legendre-Gauss-Lobatto nodes using the full hyperbolic-parabolic NSE equations. All console and file output is deactivated and the polynomial order is held at $N=9$. Each case is run three times and the performance metrics are averaged. For the strong scaling experiments, the wall time on the smallest number of GPUs in a given sweep is used as the baseline for calculations of parallel speedup. The single node wall time is used to calculate parallel efficiency in the weak scaling experiments.

\begin{table}[h]
  \centering
  \caption{
    Mesh resolutions for the strong and weak scaling tests of \galaexi on GPU-accelerated systems. For the strong scaling meshes, the number of DOFs for the entire domain is presented. For weak scaling the resolutions are given in DOFs/GPU.
  }
  \label{tab:gpu_meshes}
  \begin{tabular}{|l||cc|}
    \hline
    \multirow[c]{2}{*}{\thead{System}} & \multicolumn{2}{c|}{\thead{Mesh Resolution\\(in DOFs)}} \\
                                    & \thead{Strong} & \thead{Weak (per GPU)}\\
    \hline
    Hunter        & \num{5.243e9}  & \num{4.096e6} \\
    Frontier      & \num{16.780e9} & \num{1.024e6} \\
    JUPITER       & \num{8.389e9}  & \num{8.192e6} \\
    LUMI          & \num{2.097e9}  & \num{2.05e6}  \\
    Marenostrum 5 & \num{0.453e9}  & \num{1.77e6}  \\
    \hline
  \end{tabular}
\end{table}

For strong scaling, the resolution of a Cartesian mesh is held constant while the number of GPUs used to compute the case is increased by doubling the previous number of GPUs. For weak scaling, the mesh sizes are increased in conjunction with the number of GPUs to maintain the same load per GPU in each case. The selected mesh resolutions for each system are given in Table~\ref{tab:gpu_meshes}. All meshes for this scaling study were generated using the open-source pre-processing framework pyHOPE~\cite{kopper2025pyhope}. The final results of the scaling tests for a selection of systems are found in Figure~\ref{fig:gpu_scaling}.

\galaexi demonstrated near linear strong scaling on all systems, regardless of architecture. Strong scaling up to \num{65536} AMD MI250X GCDs on Frontier was achieved. For weak scaling, \galaexi achieves above 80\% parallel efficiency on all systems for the investigated ranges. The largest weak scaling case in that study featured a mesh of 67.1 billion DOFs on \num{65536} MI250X GCDs on Frontier with a parallel efficiency of 82.6\%.

\subsection{Scaling Performance on CPU Systems}
\label{subsec:cpu_perf}

\begin{figure*}
  \centering
  \includegraphics[width=0.9\textwidth]{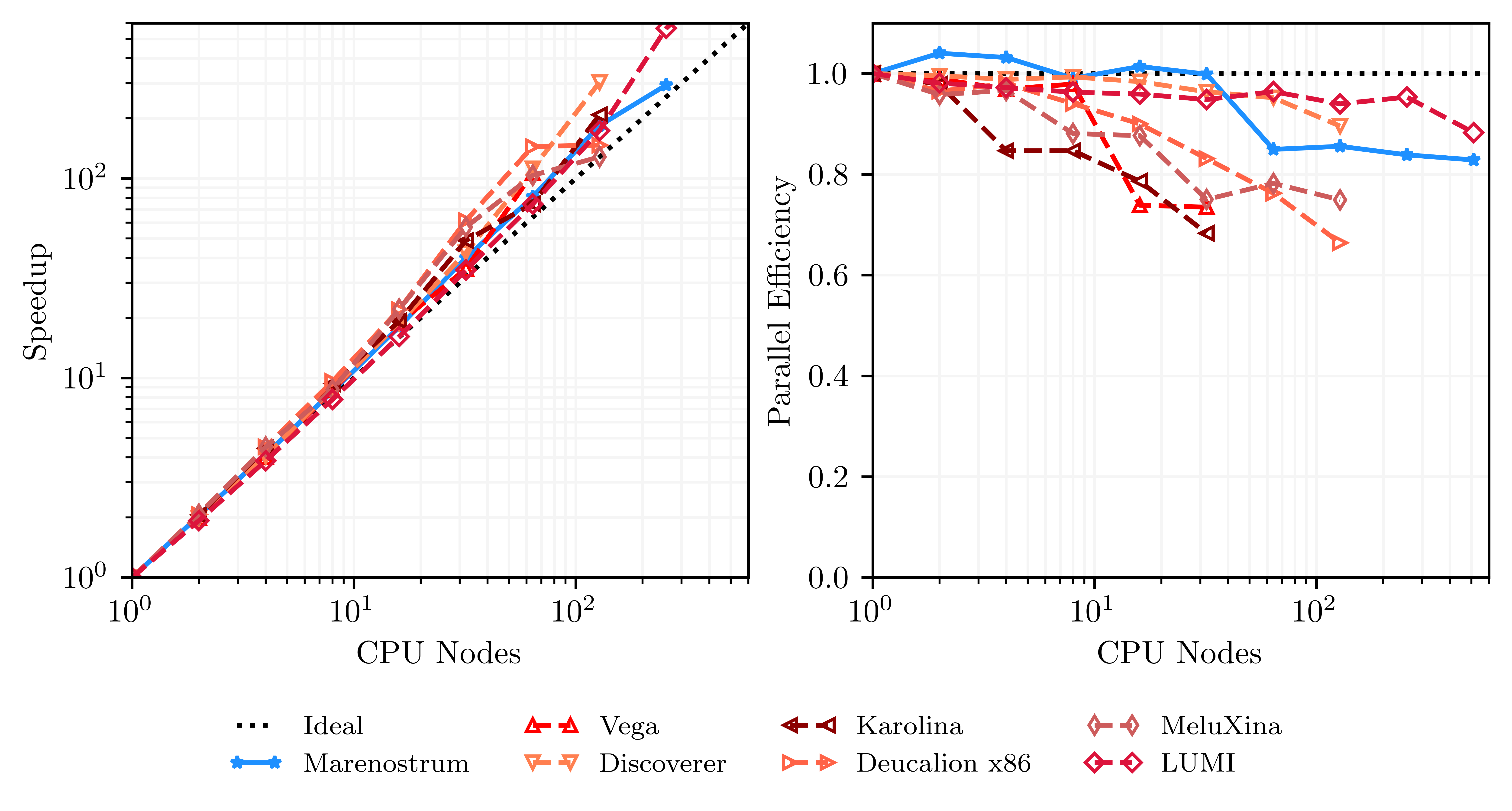}
  \caption{
    Strong (left) and weak (right) scaling performance of \galaexi on CPU-based HPC systems.
  }
  \label{fig:cpu_scaling}
\end{figure*}

\begin{table}[h]
  \centering
  \caption{
    Mesh resolutions for the strong and weak scaling tests of \galaexi on CPU systems. For the strong scaling meshes, the number of DOFs for the entire domain is presented. For weak scaling the resolutions are given in DOFs/core.
  }
  \label{tab:cpu_meshes}
  \begin{tabular}{|l||cc|}
    \hline
    \multirow[c]{2}{*}{\thead{System}} & \multicolumn{2}{c|}{\thead{Mesh Resolution\\(in DOFs)}}\\
                                    & \thead{Strong} & \thead{Weak (per core)} \\
    \hline
    Deucalion x86 & \num{2.83e6}  & \num{6192} \\
    Discoverer    & \num{2.83e6}  & \num{6192} \\
    Karolina      & \num{2.83e6}  & \num{6192} \\
    LUMI          & \num{226.5e6} & \num{6192} \\
    MeluXina      & \num{2.83e6}  & \num{6192} \\
    Marenostrum 5 & \num{56.6e6}  & \num{7899} \\
    Vega          & \num{2.83e6}  & \num{6192} \\
    \hline
  \end{tabular}
\end{table}

As was described in Section~\ref{subsec:porting}, the ability to perform CPU computations was retained in \galaexi. To measure the scaling performance of the CPU backends, scaling experiments similar to those performed for the GPU backends were repeated. For all CPU scaling tests, the polynomial order is held at $N=5$. Otherwise, the setup for the CPU scaling tests exactly mirrored those for GPUs described in Section~\ref{subsec:gpu_perf}. The mesh resolutions selected for the scaling tests on CPU systems are given in Table~\ref{tab:cpu_meshes}. The results of these experiments are shown in Figure~\ref{fig:cpu_scaling}.

Due to reduced cache pressure, \galaexi showed superlinear strong scaling on all tested systems, including on up to \num{65536} CPU cores on LUMI and \num{28672} CPU cores on Marenostrum 5. Weak scaling performance up to \num{65536} CPU cores on LUMI and \num{57344} CPU cores on Marenostrum 5 was achieved, with parallel efficiencies on those two systems higher than 80\% for the largest cases. Parallel efficiencies higher than 60\% were observed for all systems for the largest tested CPU core counts.

\subsection{Single Node Performance}
\label{subsec:sngl_node}

\begin{table*}[h]
  \centering
  \caption{
    Comparison of PID and time-to-solution on Karolina and Hunter with a problem size of 16.78 million DOFs with a polynomial order of $N=3$. Results are compared full node to full node, i.e. 128 AMD EPYC 7H12 CPU cores for Karolina CPU and 4 AMD MI300As on Hunter. \emph{\% FV} denotes the percentage of total elements switched to use FV sub-cells.
  }
  \label{tab:perf_comp}
  \begin{tabular}{|l"l|c|c|c|c|c|}
    \hline
    \thead{System} & \thead{Hardware} & \thead{\% FV} & \thead{PID [s]} & \thead{Speedup\\vs CPU} & \thead{Time To\\Solution [s]} & \thead{Speedup\\vs CPU}\\
    \Xhline{2\arrayrulewidth}
    \multirow{2}{*}{Karolina} & \multirow{2}{*}{CPU}
                              & 0 & \numThreePlaces{1.852391e-6} & -- & \num{525.13} & -- \\
                              \cline{3-7}
                              & & 50 & \numThreePlaces{3.28564e-6} & -- & \num{929.47} & -- \\
    \hline
    \multirow{2}{*}{Hunter} & \multirow{2}{*}{AMD MI300A}
                              & 0 & \numThreePlaces{3.56864e-9} & 519.1x & \num{67.73} & 7.75x \\
                              \cline{3-7}
                              & & 50 & \numThreePlaces{8.35923e-9} & 393.1x & \num{115.03} & 8.08x \\
    \hline
  \end{tabular}
\end{table*}

Table~\ref{tab:perf_comp} presents a comparison of the CPU and GPU backends with regards to both PID and time-to-solution.
For this comparison, a problem size of 16.78 million DOFs was used, with polynomial order of $N=3$. A freestream flow was simulated on a full CPU node on Karolina, which consist of 128 AMD EPYC 7H12 CPU cores~\cite{karolina}, and a full node on Hunter, with four AMD MI300A accelerated processing units \\(APUs)~\cite{hunter}. This was repeated three times and the results averaged for each architecture. Legendre-Gauss nodes were used. For the simulations featuring FV sub-cell shock capturing, 50\% of the cells are forced into the FV representation in a checkerboard pattern by a special troubled cell indicator. The simulation was allowed to run for ten timesteps and the PID and time-to-solution measured.

Comparing the performance of the two backends of \galaexi on Hunter, a 519.1x speedup in PID is seen for full DG and a 393.1x for DG/FV hybrids simulations. The theoretical throughput of an AMD MI300A for 64-bit floating point operations is \num{61e12} FLOPs~\cite{mi300a} and the corresponding value for a single AMD EPYC 7H12 core is \num{26.4e9} FLOPs~\cite{epyc_datasheet}. As the PID is effectively comparing a single AMD MI300A APU to a single CPU core, such a large speedup in PID is expected. A fairer comparison of the performance of the two architectures is the total time-to-solution. The time-to-solution measurements include serializations, such as solution initialization, data analysis and file output, which are not accounted for in the PID and are largely carried out on the CPU in \galaexi. Speedups in time-to-solution for Hunter versus the CPU computations are 7.75x and 8.08x for full DG and DG/FV hybrid, respectively, providing a more realistic picture of the benefits in real applications from GPU-acceleration in \galaexi.

\begin{table}[h]
  \centering
  \caption{
    Comparison of energy-to-solution on Marenostrum 5 and LUMI with a problem size of 16.78 million DOFs with a polynomial order of $N=7$. energy-to-solution results are compared full node to full node, i.e 112 Intel Xeon CPU cores to four NVIDIA H100s for Marenostrum 5~\cite{marenostrum} and 128 AMD EPYC cores to eight AMD MI250X GCDs on LUMI~\cite{lumi}.
  }
  \label{tab:energy_comp}
  \begin{tabular}{|l"l||c|c|}
    \hline
    \thead{System} & \thead{Hardware} & \thead{Energy To\\Solution [MJ]} & \thead{Improvement\\vs CPU} \\
    \hline
    \multirow{2}{*}{Marenostrum 5} & CPU & \num{32.16} & -- \\
                                    \cline{2-4}
                                    & GPU & \num{15.67} & 2.05x \\
                                    \cline{2-4}
    \hline
    \multirow{2}{*}{LUMI} & CPU & \num{27.43} & -- \\
                          \cline{2-4}
                          & GPU & \num{13.34} & 2.06x \\
                          \cline{2-4}
    \hline
  \end{tabular}
\end{table}

Superior throughput also enables GPUs to compute the same problem sizes on a smaller hardware footprint, improving energy consumption versus CPU-based computations. Table~\ref{tab:energy_comp} presents a comparison of energy-to-solution from the CPU and GPU backends. To collect the data in Table~\ref{tab:energy_comp}, a freestream flow was calculated on a Cartesian mesh with 16.78 million DOFs at a polynomial order of $N=7$. The simulation was allowed to progress for 1000 timesteps and energy measured on the two systems using the methods described in Section~\ref{subsec:perf_metrics}. The results in Table~\ref{tab:energy_comp} were collected on a single, full node on each system and hardware. On LUMI, this corresponds to 128 AMD EPYC 7763 CPU cores for the CPU simulations and eight AMD MI250X GCDs for the GPU simulations~\cite{lumi}. On Marenostrum 5 a single CPU node contains 112 Intel Sapphire Rapids 8480+ cores and a GPU-accelerated node features four NVIDIA H100 GPUs~\cite{marenostrum}.

The reductions in energy-to-solution for the GPU computations are 2.05x for Marenostrum 5 and 2.06x on LUMI. The GPU-accelerated simulations in \galaexi therefore require half of the energy per node as those on CPU-based systems to compute the same problem. This value also includes the energy consumed performing initialization, mesh handling and other non-compute intensive operations that are performed on the CPU even for GPU-accelerated runs.

\section{Applications}
\label{sec:application}

To demonstrate the capabilities of \galaexi for production-scale cases, it is applied to two WRLES of transonic flow over two airfoil configurations, the National Advisory Committee for Aeronautics (NACA) 64A-110 and the Office National \\d'etudes et de Recherches Aerospatiales (ONERA) OAT15A. In this section, results are presented for both configurations and the performance of the CPU and GPU backends in \galaexi are compared.

\subsection{Transonic flow over a NACA 64A-110 airfoil}
\label{subsec:naca_airfoil}

\subsubsection{Problem Description and Setup}
\label{subsubsec:naca_description}

The NACA 64A-110 was selected as an example geometry for airfoils commonly used in the horizontal tails of aircraft. From a computational perspective, it serves as a model case for compressible, turbulent flow fields \emph{without} the presence of shock waves.

The two-dimensional airfoil profile is extended to a spanwise extent of $0.05c$, where $c$ is the airfoil chord length. The airfoil is subjected to a freestream flow with a Reynolds number of $\ppRe_c=\num{9.3e5}$ based on the chord length and a Mach number of $\ppMa=0.72$. The airfoil is oriented at an angle of attack (AoA) of $\alpha=0^{\circ}$. The entropy-stable DGSEM on Legendre--Gauss--Lobatto nodes are integrated in time by using the RK4-14 for additional numerical stability. The polynomial order is set to $N=7$, resulting in a computational domain with 663 million DOFs. The mesh uses curved boundaries and nearfield volume elements, represented by a polynomial degree of $N_{Geo}=4$, to accurately represent the curvature of the airfoil geometry. To force the turbulent transition of the boundary layer, a geometric trip is placed at $x/c=0.05$. The trip is applied on both the pressure and suction sides of the airfoil. It is extended into the boundary layer to an approximate non-dimensional wall distance of $y^{+}\in[40,60]$. To prevent non-physical behavior at the farfield boundaries, an exponentially decaying filter by Pruett~\cite{pruett2003sponge} in a region starting at a radius of $50c$ from the airfoil is added. All computations were carried out on LUMI. The CPU-based simulations were performed on 256 nodes (\num{32768} CPU cores) resulting in \num{20248} DOFs per CPU core. For the GPU-accelerated case, the AMD HIP backend of \galaexi was deployed on 128 nodes (\num{1024} AMD MI250X GCDs) resulting in a load of \num{647928} DOFs/GPU.

For both architectures, benchmark simulations were initialized with a pre-converged flow state that was advanced for 67 characteristic time units (CTUs) $t^{*}=tu_{\infty}/c$, where $t$ is the solution time and $u_{\infty}$ is the freestream velocity. From this initial point, the solution is advanced in all cases for two additional CTUs. The flow data is analyzed and written to file every $0.2t^{*}$, a configuration intended to mimic production scenarios. Data analysis and output at each checkpoint includes the full instantaneous and time-averaged flow field and the local instantaneous flow state at a set of probe points along the airfoil surface.

\subsubsection{Results}
\label{subsubsec:naca_results}

\begin{figure*}
    \centering
    \includegraphics[width=0.9\textwidth]{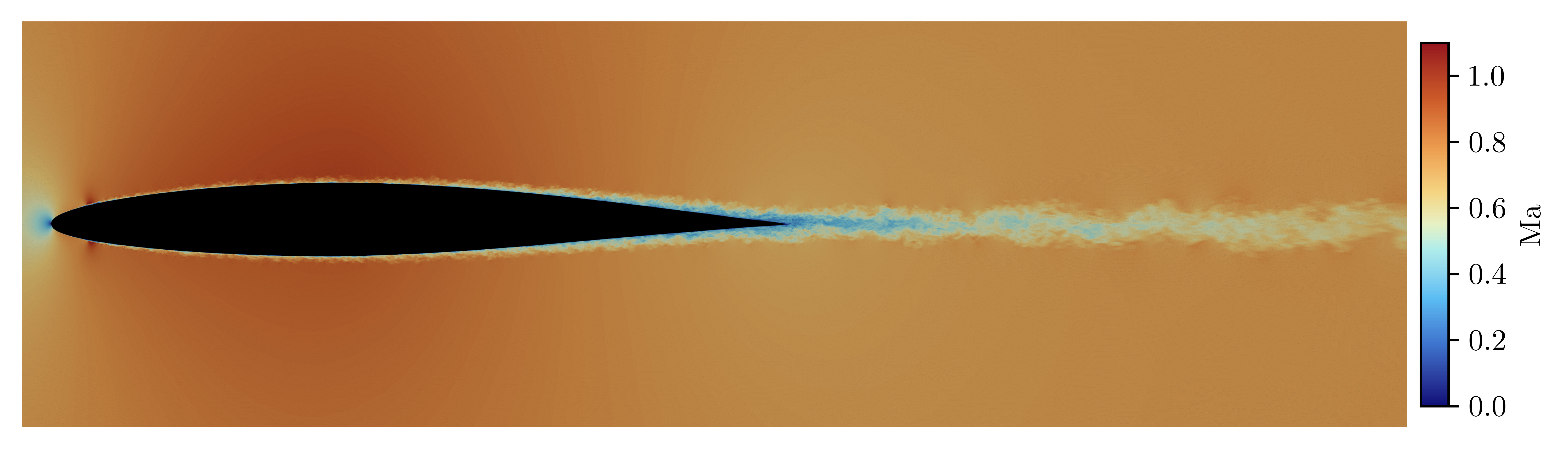}
    \caption{
        Isosurface of the instantaneous Mach number at $t^{*}=69$ for the NACA 64A-110 airfoil at $\alpha=0^{\circ}$. The visualized plane is located at the center point of the airfoil extent in the spanwise direction. Presented data was computed on LUMI with AMD MI250X GPUs.
    }
    \label{fig:naca_mach}
\end{figure*}

A representation of flow field at $t^{*}=69$ colored by local Mach number is shown in Figure~\ref{fig:naca_mach}. The results shown in this figure come from the GPU-accelerated simulation. The flow in the figure proceeds from left to right. The strong geometric trip on both sides of the airfoil are visible near the leading edge. After the trip, the boundary transitions to a fully turbulent state and well-developed turbulent structures can be seen in the immediate vicinity of the trailing edge.

\begin{figure}
    \centering
    \includegraphics[width=0.9\columnwidth]{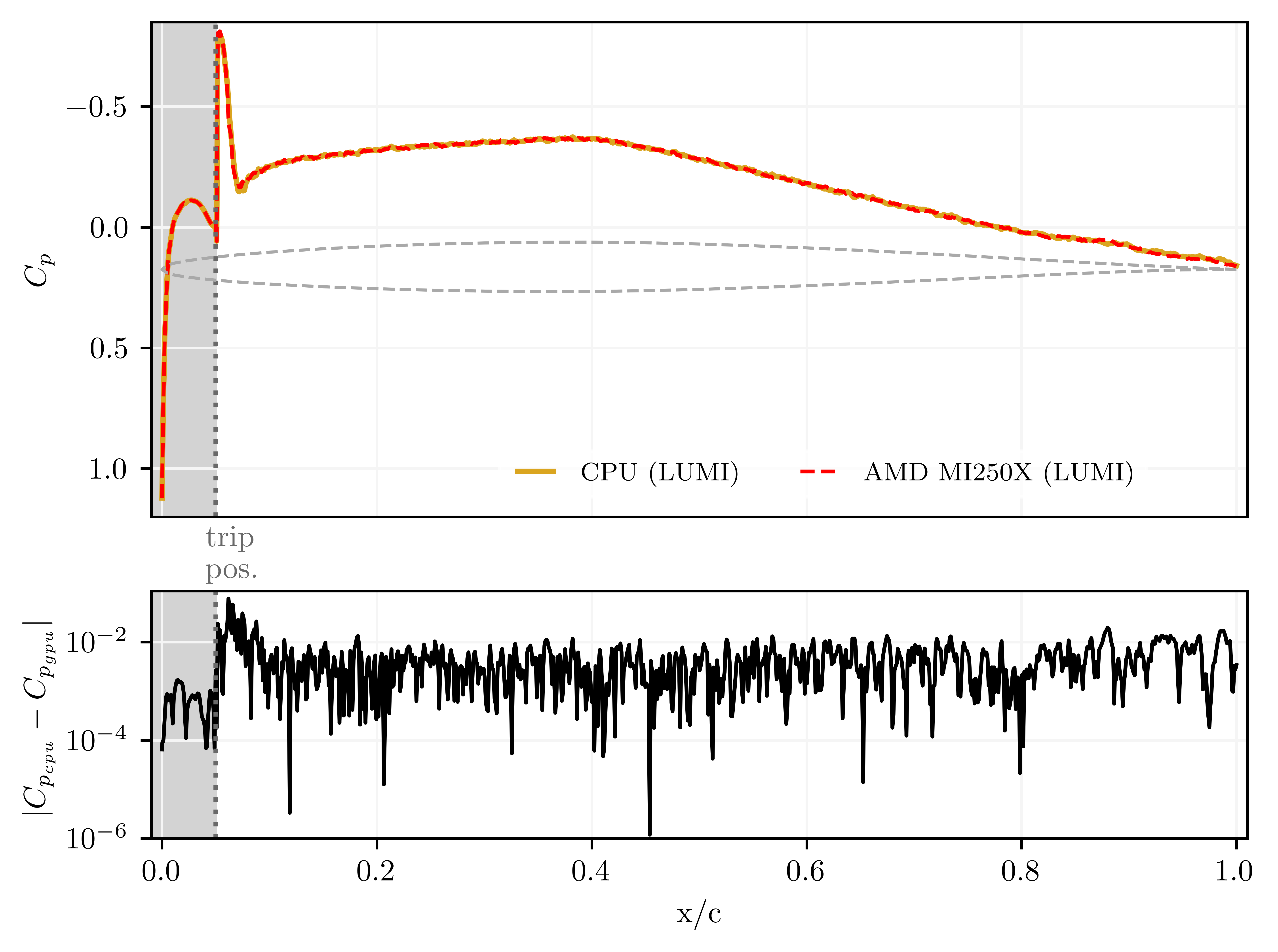}
    \caption{
      Comparison of the results for the CPU and AMD GPU backends of \galaexi for the NACA 64A-110 using the variation of surface pressure coefficient (top) along the chord. Calculation of the coefficients used spanwise-averaged flow state data at $t^{*}=69$ using data from probe points along both sides of the airfoil. The absolute difference between the two solutions (bottom) is calculated as the absolute value of the difference between the CPU and GPU-accelerated solutions.
    }
  \label{fig:naca_cp_cf}
\end{figure}

Figure~\ref{fig:naca_cp_cf} compares the evolution of the surface pressure $C_p$ coefficient along the chord of the airfoil from the CPU and AMD GPU backends of \galaexi. Spanwise-averaged flow data from the surface of the airfoil at $t^{*}=69$ is used to calculate the coefficients. The bottom plot of Figure~\ref{fig:naca_cp_cf} presents the absolute difference of the two solutions which is greatest immediately downstream of the geometric trip and settles into a range of \num{1e-4} to \num{1e-2} along the airfoil. The average percent difference along the chord length between the two architectures is 0.41\%, showing excellent agreement. Execution of both CPU and GPU backends in \galaexi is non-deterministic, so small differences in the solutions are expected. These results demonstrate \galaexi's ability to deliver robust solutions in production-scale turbulent flow cases regardless of the compute backend used.

\begin{table}[h]
  \centering
  \caption{
    Performance comparisons for simulations of the NACA 64A-110 with the CPU and AMD HIP backends of \galaexi. Computations were performed on LUMI. Note that the value of 'Ranks' for the GPU results corresponds to the number of AMD MI250X GCDs.
  }
  \label{tab:naca_perf}
  \resizebox{\columnwidth}{!}{
    \begin{tabular}{|l"c|c|c|c|}
        \hline
        \thead{Hardware} & Ranks & DOFs/Rank & \thead{Time To\\Solution\\per CTU [s]} & \thead{Speedup\\vs CPU} \\
        \Xhline{2\arrayrulewidth}
        CPU & \num{32768} & \num{20248} &\num{49097} & -- \\
        \hline
        AMD MI250X & \num{1024} & \num{647928} & \num{14172} & 3.46x \\
        \hline
    \end{tabular}
  }
\end{table}

A comparison of the achieved time-to-solution of a single CTU for all three simulations is found in Table~\ref{tab:naca_perf}. The GPU-accelerated AMD HIP backends of \galaexi required \num{3.93} hours to complete a single CTU versus \num{13.6} hours for the CPU backends, providing a improvement in time-to-solution of 3.46x.

\subsection{Transonic shock-buffet on a OAT15A airfoil}
\label{sec:oat_airfoil}

\subsubsection{Problem Description and Setup}
\label{subsubsec:oat_description}

The second application presented is an ONERA OAT15A airfoil under shock-buffet conditions. The OAT15A is a supercritical airfoil that represents a typical main wing cross-section on transport aircraft and was designed as a validation case for CFD codes~\cite{agard1994oat}. In the context of this work, the OAT15A builds upon the previous section by featuring a flow field \emph{with} shock waves and shock-turbulence interaction.

The initial and freestream flow conditions are described by a Reynolds number of $\ppRe_c=\num{3e6}$ based on the chord length of the airfoil and a Mach number of $\ppMa=0.73$. The airfoil is oriented with an AoA of $\alpha=3.5^{\circ}$ relative to the freestream. The influence of artificial reflections at the farfield boundary is prevented by choosing a large farfield radius of $80c$ and by again applying the exponential decaying Pruett filter in a region starting at a radius of $50c$ from the airfoil.

Again, the entropy-stable DGSEM on \\Legendre--Gauss--Lobatto nodes in combination with the RK4-14 for time integration is employed with a polynomial order of $N=3$. To preserve the high-order accuracy of the DGSEM, the hybrid structured-unstructured computational mesh features curved surface elements, represented by a polynomial order of $N_{Geo}=3$. Numerical stability in the presence of discontinuities is ensured by the FV sub-cell shock capturing scheme using the troubled cell indicator of Persson~\cite{persson2006Indicator}.

CPU computations with \galaexi were run with \num{25600} CPU cores (\num{200} nodes) on LUMI. This results in a load of ${\sim}\num{8000}$ DOFs/core for CPU core. GPU computations were performed on 112 NVIDIA A100 GPUs on Karolina and 112 AMD MI300A APUs on Hunter. The number of GPUs for the Karolina simulation is chosen such that the resulting computational load per GPU is ${\sim}\num{2e6}$ DOFs. For the AMD MI300A APUs featured on Hunter, the same approximate load of ${\sim}\num{2e6}$ DOFs/APU is used.

For this case, the solution for each architecture is computed separately starting from $t=0$ with the same, constant initial solution. The solution is advanced to $t^{*}=28$ with $N=1$ to overcome non-physical transient effects early in the solution. The simulations are then restarted using the same mesh and problem conditions, but using the final $N=3$ polynomial order. The solution is subsequently advanced with $N=3$ for a single buffet cycle, which corresponds to 14 CTUs, to record data for each architecture.

\subsubsection{Results}
\label{subsubsec:oat_results}
\begin{figure*}
  \centering
  \includegraphics[width=0.9\textwidth]{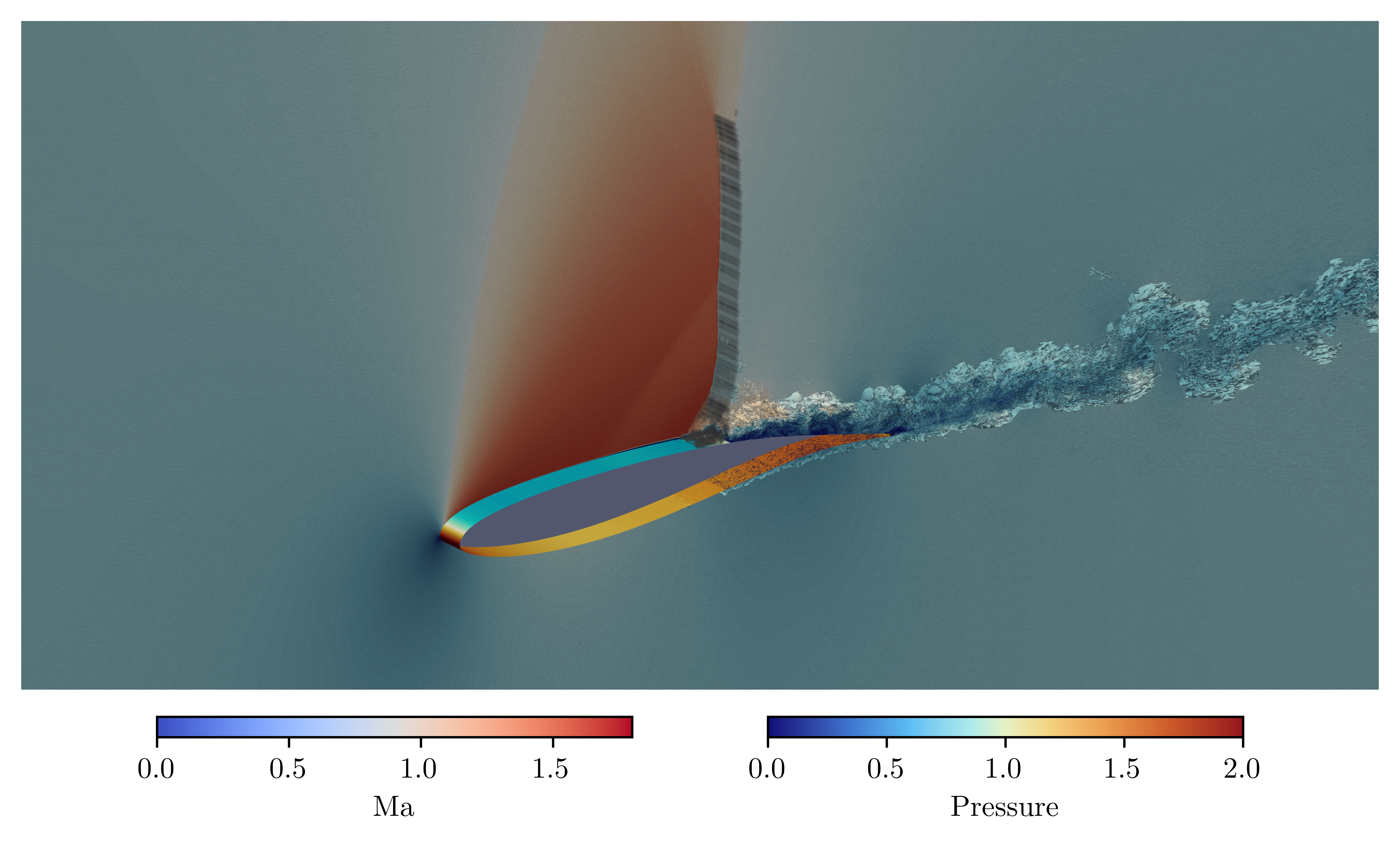}
  \caption{
      Visualization of the flow around ONERA OAT15A airfoil at $\alpha=3.5^{\circ}$ at $t^{*}=42$. The visualized plane in the background shows the flow field colored by instantaneous Mach number. The surface of the airfoil is colored by static pressure. The separated wake is demonstrated with a isosurface of Q-criterion, also colored by Mach number. Cells which have been switched to a FV representation near the shock wave are shown by the translucent grey region on the upper surface of the airfoil. Presented data was computed on Hunter with AMD MI300A APUs.
  }
  \label{fig:oat_mach}
\end{figure*}

A rendering of the instantaneous flow field at $t^{*}=42$ for the simulation performed on Hunter can be found in Figure~\ref{fig:oat_mach}. A slice plane, which shows the flow field colored by the local Mach number, forms the background of the figure. The surface of the airfoil is colored by the static pressure. The separated wake downstream of the shock wave boundary layer interaction (SWBLI) is visualized using isocontours of the Q-criterion which are colored by local Mach number. In the region of the shock wave, gray volume shading is added to those cells which have been marked by the troubled cell indicator and switched to the sub-cell FV representation.

\begin{table*}[h]
  \centering
  \caption{
    Performance comparisons for simulations of the ONERA OAT15A with the CPU and AMD HIP backends of \galaexi. The CPU and GPU-accelerated calculations were performed on LUMI and Hunter, respectively. Note that the value of 'Ranks' for the \galaexi Hunter results corresponds to the number of full AMD MI300A APUs.
  }
  \label{tab:oat_perf}
    \begin{tabular}{|l"c|c|c|c|c|}
        \hline
        \thead{Hardware} & Ranks & DOFs / Rank & \thead{Time To\\Solution\\per CTU [s]} & \thead{Speedup\\vs CPU} & \thead{Node-hours\\per CTU} \\
        \Xhline{2\arrayrulewidth}
        CPU (LUMI) & \num{25600} & \num{8668} & \num{9647.46} & -- & 536\\
        \hline
        AMD MI300A (Hunter) & 112 & \num{1981481} & \num{10690.28} & 0.90x  & 83\\
        \hline
        NVIDIA A100 (Karolina) & 112 & \num{1981481} & \num{10790.97} & 0.89x & 84\\
        \hline
    \end{tabular}
\end{table*}

A comparison of the time-to-solution and compute resources consumed for a single CTU (in node-hours) for the OAT15A across multiple architectures is given in Table~\ref{tab:oat_perf}. The GPU-accelerated AMD HIP backends of \galaexi on Hunter and the NVIDIA CUDA backends on Karolina required comparable time-to-solution to the CPU cases. However, the AMD HIP backend of \galaexi required 6.45x fewer computing resources the CPU backends to obtain the same solution in the same wall time. That directly translates to improved energy consumption and emphasizes that the gains in energy efficiency seen for single nodes in Section~\ref{subsec:sngl_node} translate to production runs of \galaexi.

\section{Conclusion}
\label{sec:conclusion}

A significant roadblock to modern computational fluid dynamics software is the need to support constantly evolving, heterogeneous computing environments. This is especially true for software intended for the study of turbulent flows, which often demand high-fidelity simulations on the largest computing resources available. To address this challenge, the discontinuous Galerkin spectral element method framework \galaexi was presented as an architecture-agnostic toolchain to enable the study of complex, compressible, turbulent flows at extreme scales.

\galaexi offers hardware-agnostic portability, achieved by interfacing a core of Fortran source code to C++ compute kernels that are compiled as either CUDA for NVIDIA GPUs or HIP for AMD GPUs. For CPU-based systems, compute routines in Fortran are retained from a CPU-only predecessor. The portability approach employed in \galaexi provides a template for similar Fortran-based software to achieve comparable results. \galaexi has also demonstrated extreme scalability. On GPUs from multiple vendors, \galaexi achieved near ideal strong and weak scaling up to \num{65536} GPUs. In the largest weak scaling case, \galaexi computed a case with 67.1 billion DOFs on \num{65536} MI250X graphics compute devices on the Frontier supercomputer with a parallel efficiency of 82.6\%, demonstrating \galaexi's ability to effectively scale up to on tens of thousands of accelerators. On CPUs, superlinear strong scaling was seen on up to \num{65536} CPU cores. Futhermore, \galaexi allows robust, high-order accurate solutions for compressible flow problems that feature shock-turbulence interactions. The solver was validated using the compressible Taylor-Green-Vortex, where solutions for all backends of \galaexi showed excellent agreement with reference data. To demonstrate \galaexi effectiveness for production-scale problems, wall-resolved large eddy simulations of transonic flows over NACA 64A-110 and ONERA OAT15A airfoil configurations were performed. For the NACA airfoil, the simulation with the \num{1024} AMD MI250X GCDs on LUMI provided a 3.46x improvement in time-to-solution over a CPU run on \num{32768} CPU cores. Results from the CPU and AMD HIP backends of \galaexi showed excellent agreement for the NACA 64A-110, producing an average percent difference in the surface pressure coefficient along the chord length of 0.41\%.

\section*{Data Availability Statement}
The source code of \galaexi is available under the \href{https://www.gnu.org/licenses/gpl-3.0.html}{GPLv3} license at:
\begin{itemize}
  \item \url{https://github.com/flexi-framework/galaexi}
\end{itemize}

The data generated in the context of this work and instructions to reproduce them with these codes are made available under the \href{https://creativecommons.org/licenses/by/4.0/}{CC-BY~4.0} license sorted by section at:
\begin{itemize}
  \item \href{https://doi.org/10.18419/DARUS-6118}{{\tt 10.18419/DARUS-6118}} (Section~\ref{sec:implementation})
  \item \href{https://doi.org/10.18419/DARUS-6120}{{\tt 10.18419/DARUS-6120}} (Section~\ref{sec:validation})
  \item \href{https://doi.org/10.18419/DARUS-6119}{{\tt 10.18419/DARUS-6119}} (Section~\ref{sec:performance})
  \item \href{https://doi.org/10.18419/DARUS-6121}{{\tt 10.18419/DARUS-6121}} (Section~\ref{sec:application})
\end{itemize}

\section*{Declaration of competing interest}
The authors declare that they have no known competing financial interests or personal relationships that could have appeared to influence the work reported in this paper.

\section*{CRediT authorship contribution statement}
\textbf{Spencer Starr:} Conceptualization, Methodology, Software, Validation, Formal Analysis, Investigation, Data Curation, Writing - Original Draft, Visualization, Project Administration \textbf{Yannik Feldner:} Methodology, Software, Investigation, Formal Analysis, Writing - Review \& Editing, Visualization \textbf{Patrick Kopper} Conceptualization, Methodology, Software, \\Formal Analysis, Writing - Review \& Editing \textbf{Marcel Blind:} Conceptualization, Methodology, Software \textbf{Daniel Kempf:} Conceptualization, Methodology, Software, Supervision \\ \textbf{Jannik Schrempp:} Software \textbf{Felix Rodach:} Software \textbf{Andrea Beck:} Writing - Review \& Editing, Resources, Supervision, Project Administration, Funding acquisition \textbf{Anna Schwarz:} Writing - Review \& Editing, Visualization, Conceptualization, Supervision

\section*{Acknowledgments}

This work was funded by the European Union. This work has received funding from the European High Performance Computing Joint Undertaking (JU) and Sweden, Germany, Spain, Greece, and Denmark under grant agreement No 101093393.

Further funding was provided by the Deutsche Forschungsgemeinschaft DFG (German Research Foundation) in the framework of the research unit FOR 2895 (Grant No. BE 6100/3-1 and BE 6100/3-2).

The authors also thank the Friedrich und Elisabeth Boysen-Stiftung for supporting the work under grant BOY187.

We acknowledge EuroHPC Joint Undertaking for awarding us access to Vega at IZUM, Slovenia, Karolina at IT4Innovations, Czech Republic, MeluXina at LuxProvide, Luxembourg, Discoverer at SofiaTech, Bulgaria, LUMI at CSC, Finland, Leonardo at CINECA, Italy, Marenostrum5 at BSC, Spain and Deucalion at MACC, Portugal.

The authors gratefully acknowledge the Gauss Centre for Supercomputing e.V. (www.gauss-centre.eu) for funding this project by providing computing time through the John von Neumann Institute for Computing (NIC) on the GCS Supercomputer JUPITER at Jülich Supercomputing Centre (JSC) as well as the support and the computing time on "Hunter" by the Supercomputing Centre Stuttgart (HLRS) through the project \\"hpcdg".

This research used resources of the Oak Ridge Leadership Computing Facility at the Oak Ridge National Laboratory, which is supported by the Advanced Scientific Computing Research programs in the Office of Science of the U.S. Department of Energy under Contract No. DE-AC05-00OR22725.

This work was completed in part at the 2024 and 2025 \\Helmholtz GPU Hackathons, part of the Open Hackathons program. The authors would like to acknowledge \\\href{https://openacc-standard.org}{OpenACC-Standard.org}, JSC, HZDR, and HIDA for their support.

The authors would also like to acknowledge individual contributions from Marius Kurz and Johanna Potyka at AMD, \\Philipp Offenh{\"a}user from Hewlett-Packard Enterprises (HPE).

\bibliographystyle{elsarticle-num-names}
\bibliography{bibliography}

@misc{top500nov20,
  title         = {{TOP500} - {N}ovember 2020},
  howpublished  = {\url{https://www.top500.org/lists/top500/list/2020/11/}},
  year          = 2020,
  month         = nov,
  note          = {Accessed: 2026-14-01}
}

@misc{top500nov25,
  title         = {{TOP500} - {N}ovember 2025},
  howpublished  = {\url{https://www.top500.org/lists/top500/list/2025/11/}},
  year          = 2025,
  month         = nov,
  note          = {Accessed: 2026-14-01}
}

@inproceedings{herten2023gpumodels,
author = {Herten, Andreas},
title = {Many {C}ores, {M}any {M}odels: {GPU} {P}rogramming {M}odel vs. {V}endor {C}ompatibility {O}verview},
year = {2023},
publisher = {Association for Computing Machinery},
doi = {10.1145/3624062.3624178},
booktitle = {Proceedings of the SC '23 Workshops of the International Conference on High Performance Computing, Network, Storage, and Analysis},
pages = {1019-1026},
numpages = {8},
series = {SC-W '23}
}

@article{bernardini2021streams,
  title = {{STREA}m{S}: {A} high-fidelity accelerated solver for direct numerical simulation of compressible turbulent flows},
  author = {Bernardini, Matteo and Modesti, Davide and Salvadore, Francesco and Pirozzoli, Sergio},
  journal = {Computer Physics Communications},
  volume = {263},
  year = {2021},
  pages = {107906}
}

@article{salvadore2024streams,
  title = {Open{MP} offload toward the exascale using {I}ntel® {GPU} {M}ax 1550: evaluation of {STREA}m{S} compressible solver},
  author = {Francesco Salvadore and Giacomo Rossi and Srikanth Sathyanarayana and Matteo Bernardini},
  journal = {The Journal of Supercomputing},
  volume = {80},
  year = {2024},
  pages = {21094--127}
}

@misc{cudafortran,
  author = {NVIDIA},
  title = {{CUDA} {F}ortran},
  howpublished = {\url{https://developer.nvidia.com/cuda-fortran}},
  note = {Accessed: 2026-26-02},
  year = {2026}
}

@misc{openmp,
  title = {Open{MP}},
  howpublished = {\url{https://www.openmp.org/}},
  note = {Accessed: 2026-26-02},
  year = {2026}
}

@misc{openacc,
  title = {Open{ACC}},
  howpublished = {\url{https://www.openacc.org/}},
  note = {Accessed: 2026-26-02},
  year = {2026}
}

@article{jansson2024neko,
  title         = {Neko: {A} modern, portable, and scalable framework for high-fidelity computational fluid dynamics},
  author        = {Jansson, Niclas and Karp, Martin and Podobas, Artur and Markidis, Stefano and Schlatter, Philipp},
  journal       = {Computers \& Fluids},
  pages         = 106243,
  year          = 2024,
  publisher     = {Elsevier},
  pages = {106243}
}

@article{jansson2025nekoperf,
  title = {Design of {N}eko—{A} {S}calable {H}igh-{F}idelity {S}imulation {F}ramework {W}ith {E}xtensive {A}ccelerator {S}upport},
  author = {Jansson, Niclas and Karp, Martin and Wahlgren, Jacob and Markidis, Stefano and Schlatter, Philipp},
  journal = {Concurrency and Computation: Practice and Experience},
  volume = {37},
  number = {2},
  year = {2025}
}

@techreport{fischer2007nek5000,
  title         = {Nek5000},
  author        = {Fischer, Paul and Lottes, James and Tufo, Henry},
  year          = 2007,
  institution   = {Argonne National Lab, Argonne, IL (United States)},
}

@article{fischer2022nekrs,
  title         = {{NekRS}, a {GPU}-accelerated spectral element {N}avier--{S}tokes solver},
  author        = {Fischer, Paul and Kerkemeier, Stefan and Min, Misun and Lan, Yu-Hsiang and Phillips, Malachi and Rathnayake, Thilina and Merzari, Elia and Tomboulides, Ananias and Karakus, Ali and Chalmers, Noel and others},
  journal       = {Parallel Computing},
  volume        = 114,
  pages         = 102982,
  year          = 2022,
  publisher     = {Elsevier},
}

@manual{fun3d14manual,
  title = {{FUN3D} {M}anual: 14.2},
  author = {Anderson, William K. and Biedron, Robert T. and Carlson, Jan-Rene{\'e} and Derlaga, Joseph M. and Diskin, Boris and Druyor Jr., Cameron T. and et al},
  year = {2025}
}

@misc{cudatoolkit,
  title = {{CUDA} {T}oolkit},
  author = {NVIDIA},
  howpublished = {\url{https://developer.nvidia.com/cuda-toolkit}},
  note = {Accessed: 2026-11-03},
  year = {2026}
}

@misc{hiplatest,
  title = {{HIP}},
  author = {{Advanced Micro Devices}},
  howpublished = {\url{https://rocm.docs.amd.com/projects/HIP/en/latest/}},
  note = {Accessed: 2026-11-03},
  year = {2026}
}

@misc{khronossycl,
  title = {{SYCL}},
  author = {{Khronos Group}},
  note = {Accessed: 2026-11-03},
  howpublished = {\url{https://www.khronos.org/sycl/}},
  year = {2026}
}

@inproceedings{nastac2021fun3dgpu,
  title = {Implicit {T}hermochemical {N}onequilibrium {F}low {S}imulations on {U}nstructured {G}rids using {GPU}s},
  author = {Nastac, Gabriel and Walden, Aaron and Nielsen, Eric J. and Frendi, Kader},
  booktitle = {AIAA Scitech 2021 Forum},
  year = {2021},
  doi = {10.2514/6.2021-0159}
}

@inproceedings{nastac2023multiarch,
  title = {A {M}ulti-{A}rchitecture {A}pproach for {I}mplicit {C}omputational {F}luid {D}ynamics on {U}nstructured {G}rids},
  author = {Nastac, Gabriel and Walden, Aaron and Wang, Li and Nielsen, Eric J. and Liu, Yu and Opgenorth, Matthew and Orender, Jason and Zubair, Mohammad},
  booktitle = {AIAA Scitech 2023 Forum},
  year = {2023},
  doi = {10.2514/6.2023-1226}
}

@article{kurz2024galaexi,
  title         = {{GAL{\AE}XI}: {S}olving complex compressible flows with high-order discontinuous {G}alerkin methods on accelerator-based systems},
  author        = {Marius Kurz and Daniel Kempf and Marcel Blind and Patrick Kopper and Philipp Offenh\"{a}user and Anna Schwarz and Spencer Starr and Jens Keim and Andrea Beck},
  year          = {2024},
  volume        = {306},
  doi           = {10.1016/j.cpc.2024.109388},
  journal       = {Computer Physics Communications},
  publisher     = {Elsevier}
}

@article{krais2021flexi,
  title         = {{FLEXI}: {A} high order discontinuous {G}alerkin framework for hyperbolic--parabolic conservation laws},
  author        = {Krais, Nico and Beck, Andrea and Bolemann, Thomas and Frank, Hannes and Flad, David and Gassner, Gregor and Hindenlang, Florian and Hoffmann, Malte and Kuhn, Thomas and Sonntag, Matthias and Munz, Claus-Dieter},
  journal       = {Computers \& Mathematics with Applications},
  volume        = 81,
  pages         = {186--219},
  year          = 2021,
  publisher     = {Elsevier},
}

@article{beck2014high,
  title         = {High-order discontinuous {G}alerkin spectral element methods for transitional and turbulent flow simulations},
  author        = {Beck, Andrea D and Bolemann, Thomas and Flad, David and Frank, Hannes and Gassner, Gregor J and Hindenlang, Florian and Munz, Claus-Dieter},
  journal       = {International Journal for Numerical Methods in Fluids},
  volume        = 76,
  number        = 8,
  pages         = {522--548},
  year          = 2014,
  publisher     = {Wiley Online Library},
}

@article{Schwarz2025,
  title = {Comparison of {E}ntropy {S}table {C}ollocation {H}igh-{O}rder {{DG}} {M}ethods for {C}ompressible {T}urbulent {F}lows},
  author = {Schwarz, Anna and Kempf, Daniel and Keim, Jens and Kopper, Patrick and Rohde, Christian and Beck, Andrea},
  year = 2025,
  journal = {Computers \& Fluids},
  volume = {303},
  pages = {106874},
  doi = {10.1016/j.compfluid.2025.106874},
}

@inbook{persson2006Indicator,
  author = {Per-Olof Persson and Jaime Peraire},
  title = {Sub-{C}ell {S}hock {C}apturing for {D}iscontinuous {G}alerkin {M}ethods},
  booktitle = {44th AIAA Aerospace Sciences Meeting and Exhibit},
  year = {2006},
  doi = {10.2514/6.2006-112}
}

@inproceedings{sonntag2014shock,
  title         = {Shock capturing for discontinuous {G}alerkin methods using finite volume subcells},
  author        = {Sonntag, Matthias and Munz, Claus-Dieter},
  booktitle     = {Finite Volumes for Complex Applications VII-Elliptic, Parabolic and Hyperbolic Problems: FVCA 7, Berlin, June 2014},
  pages         = {945--953},
  year          = 2014,
  organization  = {Springer},
}

@article{kopper2025pyhope,
  title = {{PyHOPE}: {A} {P}ython {T}oolkit for {T}hree-{D}imensional {U}nstructured {H}igh-{O}rder {M}eshes},
  journal = {Journal of Open Source Software},
  author = {Kopper, Patrick and Blind, Marcel P. and Schwarz, Anna and Kurz, Marius and Rodach, Felix and Copplestone, Stephen M. and Beck, Andrea D.},
  doi = {10.21105/joss.08769},
  url = {https://doi.org/10.21105/joss.08769},
  year = {2025},
  publisher = {The Open Journal},
  volume = {10},
  number = {115},
  pages = {8769}
}

@article{bak2022openmp,
  title = {Open{MP} application experiences: {P}orting to accelerated nodes},
  author = {Seonmyeong Bak and Colleen Bertoni and Swen Boehm and Reuben Budiardja and Barbara M. Chapman and Johannes Doerfert and Markus Eisenbach and Hal Finkel and Oscar Hernandez and Joseph Huber and Shintaro Iwasaki and Vivek Kale and Paul R.C. Kent and JaeHyuk Kwack and Meifeng Lin and Piotr Luszczek and Ye Luo and Buu Pham and Swaroop Pophale and Kiran Ravikumar and Vivek Sarkar and Thomas Scogland and Shilei Tian and P.K. Yeung},
  journal = {Parallel Computing},
  volume = {109},
  pages = {102856},
  year = {2022},
  doi = {10.1016/j.parco.2021.102856}
}

@article{trott2022kokkos,
  title={Kokkos 3: {P}rogramming {M}odel {E}xtensions for the {E}xascale {E}ra},
  author={Trott, Christian R. and Lebrun-Grandié, Damien and Arndt, Daniel and Ciesko, Jan and Dang, Vinh and Ellingwood, Nathan and Gayatri, Rahulkumar and Harvey, Evan and Hollman, Daisy S. and Ibanez, Dan and Liber, Nevin and Madsen, Jonathan and Miles, Jeff and Poliakoff, David and Powell, Amy and Rajamanickam, Sivasankaran and Simberg, Mikael and Sunderland, Dan and Turcksin, Bruno and Wilke, Jeremiah},
  journal={IEEE Transactions on Parallel and Distributed Systems},
  year={2022},
  volume={33},
  number={4},
  pages={805-817},
  doi={10.1109/TPDS.2021.3097283}
}

@inproceedings{zenker2016alpaka,
  author={Zenker, Erik and Worpitz, Benjamin and Widera, René and Huebl, Axel and Juckeland, Guido and Knüpfer, Andreas and Nagel, Wolfgang E. and Bussmann, Michael},
  booktitle={2016 IEEE International Parallel and Distributed Processing Symposium Workshops (IPDPSW)},
  title={Alpaka -- {A}n {A}bstraction {L}ibrary for {P}arallel {K}ernel {A}cceleration},
  year={2016},
  pages={631-640},
  doi={10.1109/IPDPSW.2016.50}
}

@misc{raja,
  title = {{RAJA} {P}ortability {S}uite: {E}nabling {P}erformance {P}ortable {CPU} and {GPU} {HPC} {A}pplications},
  author = {{Lawrence Livermore National Laboratory}},
  howpublished = {\url{https://computing.llnl.gov/projects/raja-managing-application-portability-next-generation\allowbreak-platforms}},
  year = {2026}
}

@misc{solanki2025adaptivecpp,
  author = {Solanki, Marco},
  title = {{A}daptive{C}pp design and architecture},
  howpublished = {\url{https://github.com/AdaptiveCpp/AdaptiveCpp/blob/develop/doc/architecture.md}},
  note = {Accessed: 2026-09-03},
  year = {2025}
}

@Inbook{blind2023towards,
  author={Blind, Marcel and Gao, Min and Kempf, Daniel and Kopper, Patrick and Kurz, Marius and Schwarz, Anna and Beck, Andrea},
  title={{Towards {E}xascale {CFD} {S}imulations {U}sing the {D}iscontinuous {G}alerkin {S}olver {FLEXI}}},
  bookTitle={High Performance Computing in Science and Engineering '23: Transactions of the High Performance Computing Center, Stuttgart (HLRS) 2023},
  year={2026},
  publisher={Springer Nature Switzerland},
  pages={207--221},
  doi={10.1007/978-3-031-91312-9_15}
}

@inproceedings{malaya2023exascale,
  author = {Bauman, Paul and Messer, Bronson and Glenski, Joseph and Georgiadou, Antigoni and Lietz, Justin and Gottiparthi, Kalyana and Day, Marc and Chen, Jackie and Rood, Jon and Esclapez, Lucas and White III, James and Jansen, Gustav R. and Curtis, Nicholas and Nichols, Stephen and Kurzak, Jakub and Chalmers, Noel and Freitag, Chip and Malaya, Nicholas and Fanfarillo, Alessandro and Budiardja, Reuben D. and Papatheodore, Thomas and Frontiere, Nicholas and Mcdougall, Damon and Norman, Matthew and Sreepathi, Sarat and Roth, Philip and Bykov, Dmytro and Wolfe, Noah and Mullowney, Paul and Eisenbach, Markus and Henry De Frahan, Marc T. and Joubert, Wayne},
  title = {Experiences readying applications for {E}xascale},
  year = {2023},
  publisher = {Association for Computing Machinery},
  doi = {10.1145/3581784.3607065},
  booktitle = {Proceedings of the International Conference for High Performance Computing, Networking, Storage and Analysis},
  articleno = {53},
  numpages = {13}
}

@article{dai2024dgcomp,
  title = {Evaluating performance portability of five shared-memory programming models using a high-order unstructured {CFD} solver},
  journal = {Journal of Parallel and Distributed Computing},
  volume = {187},
  pages = {104831},
  year = {2024},
  doi = {j.jpdc.2023.104831},
  author = {Zhe Dai and Liang Deng and YongGang Che and Ming Li and Jian Zhang and Yueqing Wang}
}

@article{reid2007fortran,
author = {Reid, John},
title = {The new features of {F}ortran 2003},
year = {2007},
publisher = {Association for Computing Machinery},
volume = {26},
number = {1},
doi = {10.1145/1243413.1243415},
pages = {10-33},
numpages = {24}
}

@misc{mi300a,
  title = {{AMD} {CDNA}$\textsuperscript{\texttrademark}$ {A}rchitecture},
  author = {{Advanced Micro Devices}},
  howpublished = {\url{https://www.amd.com/content/dam/amd/en/documents/instinct-tech-docs/white-papers/amd-cdna-3-white-paper.pdf}},
  note = {Accessed: 2026-17-03},
  year = {2025}
}

@misc{epyc_datasheet,
  title = {{AMD} {EPYC}$\textsuperscript{\texttrademark}$ 7002 {S}eries {P}rocessors},
  author = {{Advanced Micro Devices}},
  howpublished = {\url{https://www.amd.com/content/dam/amd/en/documents/products/epyc/amd-epyc-7002-series-datasheet.pdf}},
  note = {Accessed: 2026-17-03},
  year = {2025}
}

@article{klockner2009dg,
  title = {{{N}odal discontinuous {G}alerkin methods on graphics processors}},
  journal = {Journal of Computational Physics},
  volume = {228},
  number = {21},
  pages = {7863-7882},
  year = {2009},
  doi = {j.jcp.2009.06.041},
  author = {A. Klöckner and T. Warburton and J. Bridge and J.S. Hesthaven}
}

@article{chan2016dg,
  title = {{{GPU}-accelerated discontinuous {G}alerkin methods on hybrid meshes}},
  author = {Jesse Chan and Zheng Wang and Axel Modave and Jean-Francois Remacle and T. Warburton},
  journal = {Journal of Computational Physics},
  volume = {318},
  pages = {142-168},
  year = {2016},
  doi = {https://doi.org/10.1016/j.jcp.2016.04.003}
}

@article{karakus2019dg,
  title = {{A {GPU} accelerated discontinuous {G}alerkin incompressible flow solver}},
  author = {A. Karakus and N. Chalmers and K. Świrydowicz and T. Warburton},
  journal = {Journal of Computational Physics},
  volume = {390},
  pages = {380-404},
  year = {2019},
  doi = {https://doi.org/10.1016/j.jcp.2019.04.010}
}

@misc{nsightcompute,
  title = {Nsight {C}ompute\texttrademark},
  author = {NVIDIA},
  howpublished = {\url{https://developer.nvidia.com/nsight-compute}},
  note = {Accessed: 2026-03-03},
  year = {2026}
}

@misc{deucalion,
  title = {Deucalion {U}ser {G}uide},
  author = {{Minho Advanced Computing Center (MACC)}},
  howpublished = {\url{https://docs.macc.fccn.pt/}},
  note = {Accessed: 2026-19-02},
  year = {2026}
}

@misc{discoverer,
  title = {Discoverer {HPC} Docs},
  author = {{Discoverer HPC}},
  howpublished = {\url{https://docs.discoverer.bg/index.html}},
  note = {Accessed: 2026-19-02},
  year = {2026}
}

@misc{frontier,
  title = {Frontier {U}ser {G}uide},
  author = {{Oak Ridge National Laboratory (ORNL) Leadership Computing Facility}},
  howpublished = {\url{https://docs.olcf.ornl.gov/systems/frontier_user_guide.html}},
  note = {Accessed: 2026-10-04},
  year = {2026}
}

@misc{hunter,
  title = {Hunter ({HPE})},
  author = {{Höchstleistungsrechenzentrum Stuttgart (HLRS)}},
  howpublished = {\url{https://kb.hlrs.de/platforms/index.php/Hunter_(HPE)}},
  note = {Accessed: 2026-19-02},
  year = {2026}
}

@misc{jupiter,
  title = {{JUPITER} - {E}xascale for {E}urope},
  author = {{Jülich Supercomputing Centre (JSC)}},
  howpublished = {\url{https://www.fz-juelich.de/en/jsc/jupiter}},
  note = {Accessed: 2026-12-03},
  year = {2026}
}

@misc{karolina,
  title = {Karolina},
  author = {{IT4Innovations}},
  howpublished = {\url{https://www.it4i.cz/en/infrastructure/karolina}},
  note = {Accessed: 2026-19-02},
  year = {2026}
}

@misc{leonardo,
  title = {Leonardo},
  author = {{CINECA HPC}},
  howpublished = {\url{https://www.hpc.cineca.it/systems/hardware/leonardo/}},
  note = {Accessed: 2026-19-02},
  year = {2026}
}

@misc{lumi,
  title = {{LUMI} {D}ocumentation},
  author = {{CSC - IT Center for Science}},
  howpublished = {\url{https://docs.lumi-supercomputer.eu/firststeps/}},
  note = {Accessed: 2026-19-02},
  year = {2026}
}

@misc{marenostrum,
  title = {Mare{N}ostrum 5 {T}echnical information},
  author = {{Barcelona Supercomputing Center (BSC)}},
  howpublished = {\url{https://www.bsc.es/marenostrum/marenostrum-5}},
  note = {Accessed: 2026-19-02},
  year = {2026}
}

@misc{meluxina,
  title = {Melu{X}ina},
  author = {{LuxProvide}},
  howpublished = {\url{https://www.luxprovide.lu/meluxina/}},
  note = {Accessed: 2026-19-02},
  year = {2026}
}

@misc{vega,
  title = {Vega},
  author = {{Institute for Information Science, Maribor (IZUM)}},
  howpublished = {\url{https://izum.si/en/vega-en/}},
  note = {Accessed: 2026-19-02},
  year = {2026}
}

@misc{mi250x_datasheet,
  title = {{AMD} {I}nstinct$\textsuperscript{\texttrademark}$  {MI250X} {A}ccelerators},
  author = {{Advanced Micro Devices}},
  howpublished = {\url{https://www.amd.com/en/products/accelerators/instinct/mi200/mi250x.html}},
  year = {2026}
}

@misc{slurm,
  title = {Slurm {W}orkload {M}anager},
  author = {SchedMD},
  howpublished = {\url{https://slurm.schedmd.com/}},
  note = {Accessed: 2026-27-01},
  year = {2026}
}

@misc{ear,
  title = {{EAR}: {E}nergy management framework for {HPC}},
  author = {{Barcelona Supercomputing Center (BSC)}},
  howpublished = {\url{https://www.bsc.es/research-and-development/software-and-apps/software-list/ear-energy-management-framework-hpc}},
  note = {Accessed: 2026-27-01},
  year = {2026}
}

@article{Chandrashekar2013,
  title = {Kinetic {{Energy Preserving}} and {{Entropy Stable Finite Volume Schemes}} for {{Compressible Euler}} and {{Navier-Stokes Equations}}},
  author = {Chandrashekar, Praveen},
  year = 2013,
  journal = {Communications in Computational Physics},
  volume = {14},
  number = {5},
  pages = {1252--1286},
  issn = {1815-2406},
  doi = {10.4208/cicp.170712.010313a},
}

@article{Niegemann2012,
  title = {Efficient {L}ow-{S}torage {{Runge}}--{{Kutta}} {S}chemes with {O}ptimized {S}tability {R}egions},
  author = {Niegemann, Jens and Diehl, Richard and Busch, Kurt},
  year = 2012,
  journal = {Journal of Computational Physics},
  volume = {231},
  number = {2},
  pages = {364--372},
  publisher = {Elsevier Inc.},
  issn = {00219991},
  doi = {10.1016/j.jcp.2011.09.003},
}

@article{Carpenter1994,
  title = {Fourth-{{Order 2N-Storage Runge-Kutta Schemes}}},
  author = {Carpenter, Mark H and Kennedy, A},
  year = 1994,
  journal = {Nasa Technical Memorandum},
  volume = {109112},
  pages = {1--26},
}

@article{Harten1983b,
  title = {Self {A}djusting {G}rid {M}ethods for {O}ne-{D}imensional {H}yperbolic {C}onservation {L}aws},
  author = {Harten, Ami and Hyman, James M.},
  year = 1983,
  month = may,
  journal = {Journal of Computational Physics},
  volume = {50},
  number = {2},
  pages = {235--269},
  issn = {00219991},
  doi = {10.1016/0021-9991(83)90066-9},
}

@phdthesis{Sonntag2017a,
  title = {Shape {D}erivatives and {S}hock {C}apturing for the {{Navier-Stokes}} {E}quations in {D}iscontinuous {{Galerkin}} {M}ethods},
  author = {Sonntag, Matthias},
  year = 2017,
  doi = {10.18419/opus-9342},
  isbn = {978-3-8439-3281-3},
  school = {Universit\"at Stuttgart},
}

@article{sarkar1991analysis,
  title         = {The analysis and modelling of dilatational terms in compressible turbulence},
  author        = {Sarkar, Sutanu and Erlebacher, Gordon and Hussaini, M Yousuff and Kreiss, Heinz Otto},
  journal       = {Journal of Fluid Mechanics},
  volume        = 227,
  pages         = {473--493},
  year          = 1991,
  publisher     = {Cambridge University Press},
}

@techreport{agard1994oat,
  author = {Rodde, A.M. and Archambaud, J.P.},
  institution = {{Advisory Group for Aerospace Research \& Development (AGARD)}},
  title = {{OAT15A} {A}irfoil {D}ata. {AGARD} {A}dvisory {R}eport {N}um. 303 {V}olume 1: {A} {S}election of {E}xperimental {T}est {C}ases for the {V}alidation of {CFD} codes},
  year = {1994}
}

@article{pruett2003sponge,
  author = {Pruett, C. D. and Gatski, T. B. and Grosch, C. E. and Thacker, W. D.},
  journal = {Physics of Fluids},
  title = {The temporally filtered {N}avier{\textendash}{S}tokes equations: {P}roperties of the residual stress},
  year = {2003},
  number = {8},
  pages = {2127--2140},
  volume = {15},
  doi = {10.1063/1.1582858},
  publisher = {{AIP} Publishing}
}

@article{devanna2023uranos,
  title={{URANOS}: {A} {GPU} accelerated {N}avier-{S}tokes solver for compressible wall-bounded flows},
  author={De Vanna, Francesco and Avanzi, Filippo and Cogo, Michele and Sandrin, Simone and Bettencourt, Matt and Picano, Francesco and Benini, Ernesto},
  journal={Computer Physics Communications},
  volume={287},
  pages={108717},
  year={2023},
  publisher={Elsevier}
}

@article{devanna2024uranos,
  title={{URANOS}-2.0: {I}mproved performance, enhanced portability, and model extension towards exascale computing of high-speed engineering flows},
  author={De Vanna, Francesco and Baldan, Giacomo},
  journal={Computer Physics Communications},
  volume={303},
  pages={109285},
  year={2024},
  publisher={Elsevier}
}

@article{gasparino2024sod2d,
  title={{SOD2D}: {A} {GPU}-enabled spectral finite elements method for compressible scale-resolving simulations},
  author={Gasparino, Lucas and Spiga, Filippo and Lehmkuhl, Oriol},
  journal={Computer Physics Communications},
  volume={297},
  pages={109067},
  year={2024},
  publisher={Elsevier}
}

@article{witherden2025pyfr,
  title={{PyFR} v2.0.3: {T}owards industrial adoption of scale-resolving simulations},
  author={Witherden, Freddie D and Vincent, Peter E and Trojak, Will and Abe, Yoshiaki and Akbarzadeh, Amir and Akkurt, Semih and Alhawwary, Mohammad and Caros, Lidia and Dzanic, Tarik and Giangaspero, Giorgio and others},
  journal={Computer Physics Communications},
  volume={311},
  pages={109567},
  year={2025},
  publisher={Elsevier}
}

@article{dzanic2023entropy,
  title = {{P}ositivity-preserving entropy-based adaptive filtering for discontinuous spectral element methods},
  journal = {Journal of Computational Physics},
  volume = {468},
  pages = {111501},
  year = {2022},
  doi = {https://doi.org/10.1016/j.jcp.2022.111501},
  author = {T. Dzanic and F.D. Witherden}
}

@misc{mako,
  author={Mako},
  title = {Mako {T}emplates for {P}ython},
  howpublished = {\url{https://www.makotemplates.org/}},
  note = {Accessed: 2026-23-04},
  year = {2026}
}

@inproceedings{eleftherakis2025poster,
  title={{POSTER}: {P}erformance {P}ortability in {GPU}-{A}ccelerated {S}pectral {F}inite {E}lement {F}luid {S}imulations: {A} {C}ross-layer {E}xploration {A}pproach},
  author={Eleftherakis, Panagiotis-Eleftherios and Anagnostopoulos, George and Kapetanakis, Anastassis and Umair, Mohammad and Vet, Jean-Yves and Iliakis, Konstantinos and Vincent, Jonathan and Gong, Jing and Vinuesa, Ricardo and Xydis, Sotirios},
  booktitle={Proceedings of the 22nd ACM International Conference on Computing Frontiers},
  pages={228--229},
  year={2025}
}

@article{rubio2022horses3d,
  title={{HORSES3D}: a high-order discontinuous {G}alerkin solver for flow simulations and multi-physics applications},
  author={Rubio, Gonzalo},
  journal={arXiv (Cornell University)},
  year={2022}
}

@article{fujii2025scale,
  title={Scale {R}esolving {M}ethods for {A}eronautical {F}lows toward the {E}ra of “{I}ndustrial {LES}”},
  author={Fujii, Kozo and Kawai, Soshi and Gaitonde, Datta},
  journal={Flow, Turbulence and Combustion},
  volume={115},
  number={2},
  pages={405--446},
  year={2025},
  publisher={Springer}
}

@article{ceci2025grid,
  title={Grid resolution requirements for DNS of shock/boundary-layer interactions},
  author={Ceci, Alessandro and Palumbo, Andrea and Pirozzoli, Sergio},
  journal={Computers \& Fluids},
  pages={106892},
  year={2025},
  publisher={Elsevier}
}

@misc{fortranlang,
  title = {Fortran: {H}igh-performance parallel programming language},
  author = {{Fortran Community}},
  howpublished = {\url{https://fortran-lang.org/}},
  note = {Accessed: 2026-15-04},
  year = {2026}
}

@article{roache2002code,
  title         = {Code verification by the method of manufactured solutions},
  author        = {Roache, Patrick J},
  journal       = {J. Fluids Eng.},
  volume        = 124,
  number        = 1,
  pages         = {4--10},
  year          = 2002,
}

@article{Sutherland1893,
  title         = {{LII}. {T}he viscosity of gases and molecular force},
  author        = {Sutherland, William},
  year          = 1893,
  journal       = {The London, Edinburgh, and Dublin Philosophical Magazine and Journal of Science},
  volume        = 36,
  number        = 223,
  pages         = {507--531},
  doi           = {10.1080/14786449308620508},
}

@article{Lusher2021,
  title         = {Assessment of low-dissipative shock-capturing schemes for the compressible {T}aylor--{G}reen vortex},
  author        = {Lusher, David J and Sandham, Neil D},
  journal       = {AIAA Journal},
  volume        = 59,
  number        = 2,
  pages         = {533--545},
  year          = 2021,
  publisher     = {American Institute of Aeronautics and Astronautics},
}

@article{Bassi1997,
  title         = {A {H}igh-{O}rder {A}ccurate {D}iscontinuous {F}inite {E}lement {M}ethod for the {N}umerical {S}olution of the {C}ompressible {N}avier-{S}tokes {E}quations},
  author        = {Bassi, F and Rebay, S},
  year          = 1997,
  journal       = {Journal of Computational Physics},
  volume        = 131,
  number        = 2,
  pages         = {267--279},
  doi           = {10.1006/jcph.1996.5572},
  issn          = {00219991},
}

@article{Zeifang2021,
  author        = {Zeifang, Jonas and Beck, Andrea},
  doi           = {10.1016/j.jcp.2021.110475},
  journal       = {Journal of Computational Physics},
  pages         = 110475,
  title         = {{A data-driven high order sub-cell artificial viscosity for the discontinuous {G}alerkin spectral element method}},
  volume        = 441,
  year          = 2021,
}

@article{Keim2026,
  title = {Entropy {S}table {H}igh-{O}rder {D}iscontinuous {{Galerkin}} {S}pectral-{E}lement {M}ethods on {C}urvilinear, {H}ybrid {M}eshes},
  author = {Keim, Jens and Schwarz, Anna and Kopper, Patrick and Blind, Marcel and Rohde, Christian and Beck, Andrea},
  year = 2026,
  journal = {Journal of Computational Physics},
  volume = {557},
  pages = {114829},
  issn = {00219991},
  doi = {10.1016/j.jcp.2026.114829},
  urldate = {2026-03-31},
}

@article{Mossier2026,
  title = {Tackling {C}ompressible {T}urbulent {M}ulti-{C}omponent {F}lows with {D}ynamic {H}p-{A}daptation},
  author = {Mossier, Pascal and Oestringer, Philipp and J\"ons, Steven and Keim, Jens and Mavriplis, Catherine and Beck, Andrea D. and Munz, Claus-Dieter},
  year = {2026},
  journaltitle = {Computers \& Fluids},
  volume = {306},
  pages = {106928},
  doi = {10.1016/j.compfluid.2025.106928},
}

@article{kopriva1996conservative,
  title={A conservative staggered-grid Chebyshev multidomain method for compressible flows. II. A semi-structured method},
  author={Kopriva, David A},
  journal={Journal of computational physics},
  volume={128},
  number={2},
  pages={475--488},
  year={1996},
  publisher={Elsevier}
}

\end{document}